\documentclass[aps,pre,preprint,groupedaddress,amssymb]{revtex4-2}
\usepackage{graphicx} 
\usepackage{dcolumn}
\usepackage{bm}
\usepackage{amsmath}
\usepackage{epsfig}
\usepackage{color,hyperref} 
\usepackage{amssymb,dsfont}

\usepackage{appendix}

\begin{document}

\title{Ergodic properties of functionals of Gaussian processes}

\author{Vicen\c c M\'endez}
\email{vicenc.mendez@uab.cat}
\author{Carlos Herv\'as}
\author{Rosa Flaquer-Galm\'es}

\affiliation{Grup de F\'{\i}sica Estad\'{\i}stica, Departament de F\'{\i}sica. Facultat de Ci\`{e}ncies, Universitat Aut\`{o}noma de Barcelona, 08193 Barcelona, Spain}

\begin{abstract}
We derive the first two moments of generic positive stochastic functionals in terms of the one- and two-time probability density functions of the underlying random walk, and prove ergodicity of observables in stationary random walks. These general results are applied to the half-occupation time and the occupation time in an interval of a Gaussian random walk, for which we obtain exact analytic expressions for the first two moments. We then extend the analysis to scaled Brownian motion and fractional Brownian motion, computing the ergodicity breaking parameter and establishing a simple scaling form for the probability densities of occupation times. Within the framework of infinite ergodic theory, we further identify universal properties of positive observables. All analytical predictions are fully confirmed by numerical simulations.
\end{abstract}

\maketitle

\section{Introduction}
Stochastic functionals have received much attention in the last decades. They are defined as the time integration of a prescribed function of the path of a random walk up to a measurement time. The prototypical example of a stochastic functional is the so-called Brownian functional, when the random walk is Brownian. In 1949 \cite{Kac49} Kac found that the statistical properties of one-dimensional Brownian functionals can be studied by the celebrated Feynman-Kac (FK) formula \cite{Ma05}. Since then, Brownian functionals have found numerous applications across a variety of scientific fields from probability theory \cite{Kac51,Yor92} to finance \cite{Ge93}, disordered systems, and mesoscopic physics \cite{Co05}, computer science \cite{Ma05}, hydrodynamics \cite{Bau06} and meteorology \cite{MaBr02}. 

Among the most widely studied functionals are the occupation times: the occupation time in an interval and the half occupation time.
The occupation time of a random walker inside an interval is of ecological interest since it can be interpreted as the time spent by an animal within its home range. It provides useful information on the animal’s behavior when foraging. In addition, such a functional has been applied in chemical kinetics \cite{Ag84,Ag10} and magnetic systems \cite{GoLu01}.  
On the other hand, the half-occupation time or residence time, is defined as the total time spent by a walker in the positive half-space. Its distribution function for a Brownian walker is given by the seminal L\'evy’s arcsine law \cite{Le40}. Many extensions of this well known result are found in the literature. In 1958 Lamperti obtained the generalization of the arcsine law (known as Lamperti distribution) for the case of subdiffusive walkers \cite{La58}. More recently, the extension to run and tumble walkers  \cite{SiKu19} or Brownian walkers moving in heterogeneous media \cite{Si22}, have been also considered. The problem of half occupation times for Brownian walker moving in the presence of external field was considered in \cite{MaCo02,sabhapandit2006statistical} using the Kac formalism. The extension of this formalism to deal with subdiffusive walkers was addressed using the fractional FK equation \cite{Ba06,TuCaBa09,Ca10}.  
The main goal of the previously cited works is to study the statistical properties of the stochastic functionals, by finding explicitly the probability density function and moments. Among these statistical properties are the ergodic properties. 

The study of ergodic properties of certain observables related to stochastic functionals has attracted much attention \cite{WeBa09,He08,Th14,Me15,Bu10,BaFlMe23}. A widely studied measure of the ergodic behavior of the system is the ergodicity breaking parameter EB which can be found in terms of the first two moments of the stochastic functionals. This parameter is a measure for the heterogeneity among different trajectories of one ensemble, which provides useful statistical information. In general, the first and second moments can be found by solving the FK equation for the characteristic function of the functional. However, solving the FK equation is not always an easy task. It is a backward partial differential equation which can have spatial or temporal explicit dependencies that make it very difficult or even impossible to solve in many circumstances. For example, consider a Brownian walker with power-law time-depending diffusion coefficient (scaled Brownian motion). The FK equation for the functional of the scaled Brownian motion  has an explicit time-dependence multiplying the second spatial derivative which makes it impossible to get an exact analytical solution for the probability density or the moments of the functional \cite{Dh99}.   

To circumvent this problem, in this work we propose to compute the first two moments of a stochastic functional directly from the one-time and two-time probability density functions (PDFs) of the underlying random walk. In particular, if the random walk is a Gaussian process these PDFs can be expressed in terms of the autocorrelation function only. This method is especially useful to compute the first two moments and the EB parameter for the occupation times of the scaled Brownian motion and the fractional Brownian motion; two processes which have been deeply studied. However, many of their fundamental properties still remain elusive as is the case of the occupation time. 

The paper is structured as follows: In Sec. II we find general expressions for the first two moments of any positive stochastic functional from the one-time and two-time PDFs of a Gaussian random walk. The ergodic properties are introduced in Sec. III in terms of the first two moments of the functionals, where we also derive general properties for integrable observables of functionals. In Sec. IV we show that observables of stationary random walks are ergodic, i.e., their PDF converge to a Dirac delta function in the long time limit. In Sec. V we assume that the random walk is Gaussian and we obtain general expressions for the first two moments of the half occupation time and the occupation time in an interval. In Sec. VI we particularize our general results when the random walk is a Brownian motion with time-dependent diffusivity, and for the relevant cases of scaled Brownian motion and fractional Brownian motion. Our theoretical results are checked with numerical simulations and compared with previous results. Finally, the conclusions are presented in Sec. VII.

\section{The first two moments}
We apply the Kac formalism \cite{Kac51} to derive the first two moments of the PDF of a stochastic functional.
Consider the positive stochastic functional
\begin{equation}
    Z(t)=\int_{0}^{t}U[x(\tau)]d\tau
\end{equation}
where $U[x(\tau)]$ is a positive function of the stochastic trajectory $\{x(\tau); 0\leq \tau\leq t\}$ of a random walker. Let $P(Z,t)$ be the one-time PDF of $Z(t)$ and $Q(p,t)$ its characteristic function, this is, 
\begin{eqnarray}
    Q(p,t)&=&\int_{0}^{\infty}e^{-pZ}P(Z,t)dZ=\left\langle e^{-pZ(t)}\right\rangle \nonumber\\
    &=&\left\langle e^{-p\int_{0}^{t}U[x(\tau)]d\tau}\right\rangle .
    \label{mgf0}
\end{eqnarray}
Expanding the exponential in power series we have
\begin{equation}
    Q(p,t)=\left\langle \sum_{n=0}^{\infty}\frac{(-p)^{n}}{n!}Z(t)^{n}\right\rangle =\sum_{n=0}^{\infty}\frac{(-p)^{n}}{n!}\left\langle Z(t)^{n}\right\rangle .
    \label{mgf}
\end{equation}
Recall that the first moment of $P(Z,t)$ is the mean value of the functional
\begin{eqnarray}
    \left\langle Z(t)\right\rangle &=&\int_{0}^{t}\left\langle U[x(\tau)]\right\rangle d\tau\nonumber\\
    &=&\int_{0}^{t}d\tau\int_{-\infty}^{\infty}dxU(x)P(x,\tau),
    \label{Z}
\end{eqnarray}
where $P(x,t)$ is the probability of finding the walker at position $x$ at time $t$ if it was initially at $x_0$, this is, $x(\tau=0)=x_0$.
The second moment of $P(Z,t)$ is the mean square value
\begin{eqnarray}
    \left\langle Z(t)^{2}\right\rangle &=&\left\langle \left(\int_{0}^{t}U[x(\tau)]d\tau\right)^{2}\right\rangle =2!\int_{0}^{t}d\tau_{2}\int_{0}^{\tau_{2}}d\tau_{1}\left\langle U[x(\tau_{1})]U[x(\tau_{2})]\right\rangle \nonumber\\
    &=&2!\int_{0}^{t}d\tau_{2}\int_{0}^{\tau_{2}}d\tau_{1}\int_{-\infty}^{\infty}dx_{2}\int_{-\infty}^{\infty}dx_{1}U(x_{1})U(x_{2})P(x_{2},\tau_{2};x_{1},\tau_{1})
    \label{z21}
\end{eqnarray}
where $P(x_{2},\tau_{2};x_{1},\tau_{1})$ is the two-time PDF.
Proceeding analogously, for higher order moments, it can be shown that the $n$-th order moment is given by \cite{Kac51}
\begin{eqnarray}
    \left\langle Z(t)^{n}\right\rangle =n!\int_{-\infty}^{\infty}G_{n}(x,t)dx
    \label{zn}
\end{eqnarray}
where the functions $G_n(x,t)$ with $n\in \mathbb{N}$ satisfy the renewal iterative equation
\begin{eqnarray}
    G_{n+1}(x,t)=\int_{0}^{t}d\tau_1\int_{-\infty}^{\infty}dx_1U(x_1)G_{n}(x_1,\tau_1)P(x,t|x_1,\tau_1)
    \label{Gn}
\end{eqnarray}
with $P(x,t|x_1,\tau_1)$ the propagator of the process $x(t)$ and $G_0(x,t)\equiv P(x,t)$. Note that Eq. \eqref{Gn}, unlike the results in \cite{Kac51}, is general for any random walk. The propagator $P(x,t|x_1,\tau_1)$ is the probability density for the process $x(t)$ to take the value $x$ at time $t$ given that its value at time $\tau_1$ is $x_1$. In appendix \ref{app:moments} we have checked that indeed Eqs. \eqref{Z} and \eqref{z21} can be obtained from \eqref{zn} and \eqref{Gn}.
By definition of conditional probability $P(x,t;x_1,\tau_1)=P(x,t|x_1,\tau_1)P(x_1,\tau_1)$. Inserting \eqref{zn} into  \eqref{mgf}, the moment generating function $Q(p,t)$ is found as
\begin{eqnarray}
    Q(p,t)=\sum_{n=0}^{\infty}(-p)^{n}\int_{-\infty}^{\infty}G_{n}(x,t)dx.
\end{eqnarray}
Defining
\begin{eqnarray}
    Q(x,p,t)=\sum_{n=0}^{\infty}(-p)^{n}G_{n}(x,t)
\end{eqnarray}
then 
\begin{eqnarray}
   Q(p,t)=\int_{-\infty}^{\infty}Q(x,p,t)dx. 
   \label{qpt}
\end{eqnarray}
Multiplying \eqref{Gn} by $(-p)^n$ and summing over $n$ we obtain the integral equation for $Q(x,p,t)$:
\begin{eqnarray}
    Q(x,p,t)=P(x,t)-p\int_{0}^{t}d\tau_1\int_{-\infty}^{\infty}dx_1U(x_1)P(x,t|x_1,\tau_1)Q(x_1,p,\tau_1).
    \label{qxpt}
\end{eqnarray}

Once the above equation is solved for $Q(x,p,t)$, then $Q(p,t)$ follows from \eqref{qpt}. However this is usually a very difficult task.  Taking temporal and spatial derivatives of \eqref{qxpt} and using the Master equation for the propagator of the random walk one obtains a partial differential equation for $Q(x,p,t)$ called the Feynman-Kac (FK) equation. Since we are interested mainly in the first two moments, it is not necessary to find the FK equation, solve it and obtain the moments. Instead, we can find the moments directly from the propagator of the random walk using \eqref{Z} and \eqref{z21}. 
We proceed to find the moments $\left\langle Z(t)\right\rangle$ and $\left\langle Z(t)^2\right\rangle$ from the characteristic functions of the one-time and two-time PDFs. To do this, we introduce the Fourier transform of a function $f(x)$ as 
$$
\tilde{f}(k)=\mathcal{F}[f(x)]=\int_{-\infty}^{\infty}e^{ikx}f(x)dx
$$
and the inverse Fourier transform as
$$
f(x)=\mathcal{F}^{-1}[\tilde{f}(k)]=\frac{1}{2\pi}\int_{-\infty}^{\infty}e^{-ikx}\tilde{f}(k)dk.
$$
Then, the one- and two-time PDFs can be written as
$$
P(x,t)=\frac{1}{2\pi}\int_{-\infty}^{\infty}e^{-ikx}\tilde{P}(k,t)dk
$$
and
$$
P(x_{2},t_{2};x_{1},t_{1})=\frac{1}{\left(2\pi\right)^{2}}\int_{-\infty}^{\infty}dk_{1}\int_{-\infty}^{\infty}dk_{2}\tilde{P}(k_{2},t_{2};k_{1},t_{1})e^{-i(k_{1}x_{1}+k_{2}x_{2})},
$$
respectively. In consequence, Eq. \eqref{Z} becomes 
\begin{eqnarray}
    \left\langle Z(t)\right\rangle =\frac{1}{2\pi}\int_{0}^{t}d\tau_{1}\int_{-\infty}^{\infty}dk_{1}\tilde{P}(k_{1},\tau_{1})\tilde{U}(-k_{1})
    \label{z1f}
\end{eqnarray}
where
\begin{eqnarray}
  \tilde{U}(-k_{i})=\int_{-\infty}^{\infty}e^{-ik_{i}x_{i}}U(x_{i})dx_{i}
  \label{uf}
\end{eqnarray}
and $i=1,2$.
Analogously, Eq. \eqref{z21} can be expressed as
\begin{eqnarray}
    \left\langle Z(t)^{2}\right\rangle =\frac{2}{(2\pi)^{2}}\int_{0}^{t}d\tau_{2}\int_{0}^{\tau_{2}}d\tau_{1}\int_{-\infty}^{\infty}dk_{1}\int_{-\infty}^{\infty}dk_{2}\tilde{P}(k_{2},\tau_{2};k_{1},\tau_{1})\tilde{U}(-k_{1})\tilde{U}(-k_{2}).
    \label{z2f}
\end{eqnarray}
Equations \eqref{z1f} and \eqref{z2f} are general expressions which provide a method of calculation of the first and second moments of any positive stochastic functional from the one-time and two-time PDFs of any Gaussian random walk. With these two moments, we can study the ergodic properties of such functionals as we show below.

\section{Ergodicity}
The ergodic properties of observables of random walks have attracted much interest in recent years \cite{WeBa09,He08,Th14,Me15,Bu10}. These properties can be studied by constructing the ergodicity breaking (EB) parameter, which is a measure for the heterogeneity among different trajectories of one ensemble. By definition, it is zero when the observable is ergodic and non zero when it is non-ergodic. If the mean value of the observable can be linked to the moments of a stochastic functional then the EB can be expressed in terms of the first two moments of the functional as we show below.

Let us consider a stochastic trajectory $x(\tau)$  observed from $\tau=0$ up to time $\tau = t$. Consider an observable $\mathcal{O}[x(\tau)]$, a function of the trajectory $x(\tau)$. Since $x(\tau)$ is stochastic in nature, the observable $\mathcal{O}[x(\tau)]$ will also be fluctuating between the realizations. An observable of the random walk is said to be ergodic if the ensemble average equals the time average $\left\langle \mathcal{O}\right\rangle =\ensuremath{\overline{\mathcal{O}}}$ in the long time limit. This means that if $\mathcal{O}[x(\tau)]$ is ergodic, then its time average $\ensuremath{\overline{\mathcal{O}}}$ is not a random variable. As a consequence, \textcolor{black}{the limiting PDF of $\ensuremath{\overline{\mathcal{O}}}$, namely $P(\overline{\mathcal{O}},t)$ is a Dirac delta function centered on its ensemble average:
\begin{eqnarray}
P(\overline{\mathcal{O}},t\to\infty)=\delta\left(\overline{\mathcal{O}}-\left\langle \mathcal{\overline{O}}\right\rangle \right).
     \label{lpdf}
 \end{eqnarray}} 
Recall that the probability density $P(x,t)$ is the probability to find the walker at point $x$ at time $t$.
 If the observable is integrable with respect to the density $P(x,t)$, then the ensemble average is given by
\begin{equation}
    \left\langle \mathcal{O}[x(t)]\right\rangle =\int_{-\infty}^{\infty}\mathcal{O}[x]P(x,t)dx.
\end{equation}
The time average of $\mathcal{O}[x(t)]$ is defined as
\begin{equation}
    \ensuremath{\overline{\mathcal{O}[x(t)]}=}\frac{1}{t}\int_{0}^{t}\mathcal{O}[x(\tau)]d\tau.
    \label{tav}
\end{equation}
For non-ergodic observables, since $\overline{\mathcal{O}}$ is random, its variance $\textrm{Var}(\overline{\mathcal{O}})$ is non-zero in the long time limit. Otherwise, for an ergodic observable $\textrm{Var}(\overline{\mathcal{O}})=0$ in the long time limit. Keeping this in mind, the ergodicity breaking parameter EB is defined as
\begin{eqnarray}
\textrm{EB}=\lim_{t\to\infty}\frac{\textrm{Var}(\overline{\mathcal{O}})}{\left\langle \overline{\mathcal{O}}\right\rangle ^{2}}=\lim_{t\to\infty}\frac{\left\langle \overline{\mathcal{O}}^{2}\right\rangle -\left\langle \overline{\mathcal{O}}\right\rangle ^{2}}{\left\langle \overline{\mathcal{O}}\right\rangle ^{2}}.
  \label{EB}
\end{eqnarray}
For ergodic observables, one should have $\textrm{EB}= 0$. 

In the examples below we consider the observable $\mathcal{O}[x(t)]=U[x(t)]$ so that the time average of the observable is from \eqref{tav}
\begin{eqnarray}
    \ensuremath{\overline{\mathcal{O}[x(t)]}=}\frac{1}{t}\int_{0}^{t}U[x(\tau)]d\tau=\frac{Z(t)}{t},
\end{eqnarray}
and so
\begin{eqnarray}
 \ensuremath{\left\langle \overline{\mathcal{O}}\right\rangle =}\frac{\left\langle Z(t)\right\rangle }{t},\quad\ensuremath{\left\langle \overline{\mathcal{O}}^{2}\right\rangle =}\frac{\left\langle Z(t)^{2}\right\rangle }{t^{2}}.   
 \label{o}
\end{eqnarray}
Finally, from \eqref{EB} we can find the EB in terms of the first two moments of the functional
\begin{eqnarray}
    \textrm{EB}=\frac{\left\langle Z(t)^{2}\right\rangle }{\left\langle Z(t)\right\rangle ^{2}}-1
    \label{EB2}
\end{eqnarray}
as $t\to \infty$. 

\subsection{Infinite Ergodic Theory}\label{sec:infinite}
In the long time limit, a system may reach a steady state,
namely $P(x, t )$  is time independent for long times, as we consider in the next section. This solution is usually reached from most typical initial conditions, and the time-independent density is called the invariant density \cite{von41}. Let us assume that in the long time limit the one-time PDF of the process $x(t)$ can be written as $P(x,t)\simeq \phi(t)\mathcal{I}_\infty (x)$. If the integral $\int_{-\infty}^\infty \mathcal{I}_\infty (x) dx $ is divergent then $\mathcal{I}_\infty (x)$ is an infinite invariant density and clearly $P(x,t)$ is non-normalizable. In this case, a different type of ergodic framework emerges: the  so-called infinite ergodic theory.

Consider an observable $\mathcal{O}[x(t)]$, which depends on the realizations of the process $x(t)$. 
If the observable fulfills the requirement
\begin{eqnarray}
    \int_{-\infty}^{\infty}\mathcal{O}(x)\mathcal{I}_{\infty}(x)dx<\infty,
    \label{eq:integrability}
\end{eqnarray}
namely the observable is integrable with respect to $\mathcal{I}_{\infty}(x)$, then the ensemble average over the realizations of the process $x(t)$ can be represented through the infinite invariant density for long times as
\begin{eqnarray}
    \lim_{t\to\infty}\left\langle \mathcal{O}\right\rangle =\phi(t)\int_{-\infty}^{\infty}\mathcal{O}(x)\mathcal{I}_{\infty}(x)dx.
    \label{limo}
\end{eqnarray}
Now consider the ensemble average of the time average

\begin{eqnarray}
    \left\langle \overline{\mathcal{O}}\right\rangle =\int_{-\infty}^{\infty}dxP(x,t)\frac{1}{t}\int_{0}^{t}\mathcal{O}[x(t')]dt'=\frac{1}{t}\int_{0}^{t}dt'\int_{-\infty}^{\infty}\mathcal{O}(x)P(x,\textcolor{black}{t'})dx
\end{eqnarray}

so that using \eqref{limo} we find
\begin{eqnarray}
    \lim_{t\to\infty}\left\langle \overline{\mathcal{O}}\right\rangle =\frac{1}{t}\int_{0}^{t}\phi(t')dt'\int_{-\infty}^{\infty}\mathcal{O}(x)\mathcal{I}_{\infty}(x)dx.
    \label{limo2}
\end{eqnarray}
Since $\overline{\mathcal{O}}$ is a random variable determined by the
realization of $x(t)$ thus we define the stochastic variable
\begin{eqnarray}
    \xi=\lim_{t\to\infty}\frac{\overline{\mathcal{O}}}{\left\langle \mathcal{O}\right\rangle }.
    \label{xi}
\end{eqnarray}
The observable is ergodic when $\xi =1$, and so the PDF of $\xi$ is $P(\xi,t)=\delta (\xi -1)$ in the long time limit. 
Hence, using \eqref{limo} and \eqref{limo2} we conclude that
\begin{eqnarray}
    \lim_{t\to\infty}\frac{\left\langle \overline{\mathcal{O}}\right\rangle }{\left\langle \mathcal{O}\right\rangle }=\left\langle \xi\right\rangle =\frac{1}{t\phi(t)}\int_{0}^{t}\phi(t')dt'
    \label{quo}
\end{eqnarray}
namely, the ensemble average of the time average and the ensemble average itself are related. Note that this relation is general for any observable if it is integrable with respect to the infinite invariant density, and only depends on the PDF of $x(t)$. Clearly, when the observable is ergodic  $\lim_{t\to\infty}\left\langle \overline{\mathcal{O}}\right\rangle /\left\langle \mathcal{O}\right\rangle =1$. If we choose $\mathcal{O}(x)=U(x)$ then using \eqref{quo} we find
\begin{eqnarray}
    \left\langle Z(t)\right\rangle \approx\left\langle U\right\rangle \frac{\int_{0}^{t}\phi(t')dt'}{\phi(t)},\quad t\to \infty.
    \label{mZ}
\end{eqnarray}
The infinite ergodic theory also provides an expression for the long time limit of the mean value of a functional. We compare below this result with that obtained from Eq. \eqref{Z} for specific examples.

\section{Stationary processes }
In this section we derive the temporal dependence of the moments of the functional $Z(t)$ when the underlying random walk $x(t)$ is a stationary process,
which means that it is time homogeneous and that there exists a time-independent stationary PDF defined as $P_s(x)=\lim_{t\to\infty}P(x,t)=\lim_{t\to\infty}P(x,t|x',t')$. For example, when Brownian particles are confined in a finite domain, after a sufficiently long time their concentration becomes uniform (for reflecting boundary conditions) and thus time invariant.
Another example is the case of a Brownian particle under the effect of an external confining potential \cite{Ca10,Ba06,sabhapandit2006statistical}, moving through a heterogeneous media \cite{LeBa19} or under the presence of a stochastic resetting \cite{MeFlPa25}. In the long time limit \eqref{Gn} reduces to
\begin{eqnarray}
  G_{n+1}(x,t)\simeq P_{s}(x)\int_{0}^{t}d\tau_{1}\int_{-\infty}^{\infty}dx_{1}U(x_{1})G_{n}(x_{1},\tau_{1}).  
  \label{gnss}
\end{eqnarray}
Making use of the Laplace transform defined as $\mathcal{L}[f(t)]=f(s)=\int_{0}^{\infty}e^{-st}f(t)dt$, Eq. \eqref{gnss} turns into
\begin{eqnarray}
    G_{n+1}(x,s)\simeq\frac{P_{s}(x)}{s}\int_{-\infty}^{\infty}dx_{1}U(x_{1})G_{n}(x_{1},s).
    \label{gnssn}
\end{eqnarray}
The first moment in the long time limit follows from \eqref{gnssn} and setting $n=0$ to find
\begin{eqnarray}
    \left\langle Z(s)\right\rangle \simeq\frac{1}{s^{2}}\int_{-\infty}^{\infty}dxU(x)P_{s}(x),
\end{eqnarray}
which in the real space reads
\begin{eqnarray}
        \left\langle Z(t)\right\rangle \simeq t\int_{-\infty}^{\infty}dxU(x)P_{s}(x).
        \label{z1tss}
\end{eqnarray}
The second moment can be derived from \eqref{gnssn} setting $n=1$, so that
\begin{eqnarray}
    \left\langle Z(s)^{2}\right\rangle \simeq\frac{2!}{s^{3}}\left[\int_{-\infty}^{\infty}dxU(x)P_{s},(x)\right]^{2}
\end{eqnarray}
so that,
\begin{eqnarray}
       \left\langle Z(t)^{2}\right\rangle \simeq t^2\left[\int_{-\infty}^{\infty}dxU(x)P_{s}(x)\right]^{2}. 
\end{eqnarray}
In general,
\begin{eqnarray}
    \left\langle Z(s)^{n}\right\rangle \simeq\frac{n!}{s^{n+1}}\left\langle U\right\rangle _{s}^{n}
\end{eqnarray}
where $\left\langle U\right\rangle _{s}=\int_{-\infty}^{\infty}U(x)P_{s}(x)dx$. After Laplace inversion 
\begin{eqnarray}
    \left\langle Z(t)^{n}\right\rangle \simeq\left(t\left\langle U\right\rangle _{s}\right)^{n}
    \label{Zn}
\end{eqnarray}
as $t\to \infty$. This is a general result, regardless of the functional and the details of the random walk. In addition, we can make use of \eqref{mgf} to find the characteristic function of the functional in the long time limit. Using \eqref{mgf} and \eqref{Zn} one has $Q(p,t)\simeq e^{-pt\left\langle U\right\rangle _{s}}$ and Laplace inverting with respect to the variable $p$ as defined in \eqref{mgf0} one finds 
\begin{eqnarray}
P(Z,t)\simeq\delta\left(Z-t\left\langle U\right\rangle _{s}\right)=\delta\left(Z-\left\langle Z(t)\right\rangle \right)\quad \text{as}\quad t\to \infty,
    \label{Qs}
\end{eqnarray}
i.e., the ergodicity breaking parameter is $\textrm{EB}=0$, in agreement with Khinchin’s theorem \cite{Kh49}, which provides the condition that a stationary process is ergodic. The same can be inferred from \eqref{quo}. If the underlying random walk is stationary then $\lim_{t\to\infty}\left\langle U[x(t)]\right\rangle =\left\langle U\right\rangle _{s}$ and considering $\mathcal{O}[x(t)]=U[x(t)]$ one has $\left\langle \mathcal{O}[x(t)]\right\rangle \approx\left\langle U\right\rangle _{s}$. Taking the temporal average $\left\langle \overline{\mathcal{O}}[x(t)]\right\rangle \approx\frac{1}{t}\int_{0}^{t}\left\langle U\right\rangle _{s}dt=\left\langle U\right\rangle _{s}$ so that   $\lim_{t\to\infty}\left\langle \overline{\mathcal{O}}\right\rangle /\left\langle \mathcal{O}\right\rangle =1$ and the observable is ergodic.

\section{Gaussian processes}
Let us specify our general results for the first two moments of a stochastic functional to the case where the position of the random walker $x(t)$ is a Gaussian process. To find the characteristic functions of the one-time and two-time PDFs we make use of the characteristic functional of a Gaussian process. If $x(t)$ is a Gaussian process with mean $\left\langle x(t)\right\rangle$  and autocorrelation $C(t_{1},t_{2})\equiv\left\langle x(t_{1})x(t_{2})\right\rangle -\left\langle x(t_1)\right\rangle \left\langle x(t_2)\right\rangle$ the characteristic functional of $x(t)$ is the characteristic function of the functional  $\int_{-\infty}^{\infty}f(t)x(t)dt$ and is given by \cite{vK92}
\begin{eqnarray}
    \left\langle e^{i\int_{-\infty}^{\infty}f(t)x(t)dt}\right\rangle =\exp\left[i\int_{-\infty}^{\infty}f(t)\left\langle x(t)\right\rangle dt-\frac{1}{2}\int_{-\infty}^{\infty}dt_{1}\int_{-\infty}^{\infty}dt_{2}f(t_{1})f(t_{2})C(t_{1},t_{2})\right]
    \label{cf}
\end{eqnarray}
where $f(t)$ is an arbitrary function. To find the expression for the one-time and two-time PDFs we consider $f(\tau)=k\delta (\tau -t)$ and $f(\tau)=k_1\delta (\tau -t_1)+k_2\delta (\tau -t_2)$ respectively. Thus,
\begin{eqnarray}
  P(k,t)=\left\langle e^{ikx(t)}\right\rangle =\exp\left[ik\left\langle x(t)\right\rangle -\frac{k^{2}}{2}C(t,t)\right] 
  \label{cfot}
\end{eqnarray}
and
\begin{eqnarray}
    P(k_{2},t_{2};k_{1},t_{1})=\left\langle e^{ik_{1}x(t_{1})+ik_{2}x(t_{2})}\right\rangle &=&\exp\bigg[ik_{1}\left\langle x(t_{1})\right\rangle + ik_{2}\left\langle x(t_{2})\right\rangle \nonumber\\ 
    &-&\frac{k_{1}^{2}}{2}C(t_{1},t_{1})-\frac{k_{2}^{2}}{2}C(t_{2},t_{2})-k_{1}k_{2}C(t_{1},t_{2})\bigg].
     \label{cftt}
\end{eqnarray}
If we assume for simplicity that the Gaussian process is isotropic, performing the inverse Fourier transform to \eqref{cfot}, the one-time PDF is $P(x,t)=[2\pi C(t,t)]^{-1/2}\exp[-x^2/2C(t,t)]$. For $t\to \infty$ and $x\ll \sqrt{2C(t,t)}$ we have $P(x,t)\approx [2\pi C(t,t)]^{-1/2} $, so that, $\mathcal{I}_\infty (x)=1/\sqrt{2\pi}$ and $\phi(t)=[C(t,t)]^{-1/2}$. Then, from \eqref{quo} we find
\begin{eqnarray}
\lim_{t\to\infty}\frac{\left\langle \overline{\mathcal{O}}\right\rangle }{\left\langle \mathcal{O}\right\rangle }=\frac{\sqrt{C(t,t)}}{t}\int_{0}^{t}\frac{dt'}{\sqrt{C(t',t')}}.
\label{quo2}
\end{eqnarray}
In addition, from \eqref{mZ} we obtain 
\begin{eqnarray}
    \left\langle Z(t)\right\rangle \approx\frac{\mathcal{A}}{\sqrt{2\pi}}\int_{0}^{t}\frac{dt'}{\sqrt{C(t',t')}},
    \label{mZ2}
\end{eqnarray}
where $\mathcal{A}=\int_{-\infty}^{\infty}U(x)dx<\infty$.
Recall \eqref{quo2} and \eqref{mZ2} hold for any integrable observable of an isotropic Gaussian process and is consequence of the infinite ergodic theory. 

Now, we want to particularize to the case of occupation times as examples of stochastic functionals. We consider first the half occupation time $T^+(t)$. It is defined as the total amount of time that the process $x(t)$ has positive values during the time interval $[0,t]$:
$$
T^+(t)=\int_0^t\theta[x(\tau)]d\tau.
$$
Then,  we take into account that $U(x)=\theta (x)$, with $\theta (x)$ the Heaviside function: $U=1$ if $x>0$ and $U=0$ if $x<0$. Then, from \eqref{uf} (see appendix \ref{app:eq48} for details of the derivation)
\begin{eqnarray}
    \tilde{U}(-k)=-\frac{i}{k}+\pi\delta(k).
    \label{tfU1}
\end{eqnarray}

The mean half occupation time follows from Eq. \eqref{z1f},  and \eqref{tfU1} and is given by
\begin{eqnarray}
\left\langle T^{+}(t)\right\rangle =\frac{t}{2}-\frac{i}{2\pi}\int_{0}^{t}d\tau\int_{-\infty}^{\infty}\frac{\tilde{P}(k,\tau)}{k}dk.
    \label{tdd2}
\end{eqnarray}
Finally, using \eqref{cfot} we find from \eqref{tdd2}
\begin{eqnarray}
\left\langle T^{+}(t)\right\rangle =\frac{t}{2}+\frac{1}{2}\int_{0}^{t}\textrm{erf}\left(\frac{\left\langle x(\tau)\right\rangle }{\sqrt{2C(\tau,\tau)}}\right)d\tau,
    \label{meantm}
\end{eqnarray}
which is a general expression for the mean half occupation time of a Gaussian random walk in terms of $\left\langle x(t)\right\rangle $ and $\left\langle x(t)^{2}\right\rangle =C(t,t)$.
Note that for an isotropic random walk the mean position is zero and $\left\langle T^{+}(t)\right\rangle =t/2$, which means that in this case the walker spends half of the time in the positive axis in mean. 

The mean square occupation time follows from \eqref{z2f} and  \eqref{tfU1}. Performing the calculations detailed in appendix \ref{app:eqt2g}, we get
\begin{eqnarray}
\left\langle T^{+}(t)^{2}\right\rangle &=&\left\langle T^{+}(t)\right\rangle ^{2}\nonumber\\
&+&\frac{1}{\pi}\int_{0}^{t}d\tau_{2}\int_{0}^{\tau_{2}}d\tau_{1}\int_{0}^{\frac{C(\tau_{1},\tau_{2})}{\sqrt{C(\tau_{1},\tau_{1})C(\tau_{2},\tau_{2})}}}\frac{dz}{\sqrt{1-z^{2}}}\exp\left[-\frac{\xi_{1}^{2}+\xi_{2}^{2}-2z\xi_{1}\xi_{2}}{2(1-z^{2})}\right]
 \label{t2g}
\end{eqnarray}
where
$$
\xi_{i}=\frac{\left\langle x(\tau_{i})\right\rangle }{\sqrt{C(\tau_{i},\tau_{i})}}, \quad i=1,2.
$$

Next, we consider the occupation time in the interval $[-a,a]$, say $T_a(t)$ during the time interval $[0,t]$:
$$
T_a(t)=\int_0^t \mathds{1}_{[-a,a]}(x(\tau))d\tau.
$$
To this end we consider $U(x)=\mathds{1}_{[-a,a]}(x)$, being $\mathds{1}_{[-a,a]}$ the indicator function of $[-a,a]$. Hence, using Eq. \eqref{uf} we obtain the expression
\begin{eqnarray}
\tilde{U}(-k)=\frac{2\sin(ka)}{k}
    \label{tfU2}
\end{eqnarray}
which after being introduced into Eq. \eqref{z1f} and using \eqref{cfot} gives
\begin{equation}
\left\langle T_{a}(t)\right\rangle =\frac{1}{\pi}\int_{0}^{t}d\tau\int_{-\infty}^{\infty}\frac{\sin(ka)}{k}e^{ik\left\langle x(\tau)\right\rangle -\frac{k^{2}}{2}C(\tau,\tau)}dk.
\label{ta1g}
\end{equation}

Integrating over $k$ as detailed in appendix \ref{app:Ta}, we obtain the general expression for $\langle T_a(t)\rangle$ in terms of $\langle x(t) \rangle$ and autocorrelation of the Gaussian process:

\begin{eqnarray}
    \left\langle T_{a}(t)\right\rangle =\frac{1}{2}\int_{0}^{t}\left[\textrm{erf}\left(\frac{a-\left\langle x(\tau)\right\rangle }{\sqrt{2C(\tau,\tau)}}\right)+\textrm{erf}\left(\frac{a+\left\langle x(\tau)\right\rangle }{\sqrt{2C(\tau,\tau)}}\right)\right]d\tau.
    \label{Ta1}
\end{eqnarray}

Now, the second moment $\left\langle T_{a}(t)^2\right\rangle$ is found from \eqref{z2f} and can be rewritten as
\begin{eqnarray}
    \left\langle T_{a}(t)^{2}\right\rangle =\frac{2}{\pi^{2}}\int_{0}^{t}d\tau_{2}\int_{0}^{\tau_{2}}d\tau_{1}\int_{-\infty}^{\infty}dk_{1}\frac{\sin(k_{1}a)}{k_{1}}\int_{-\infty}^{\infty}dk_{2}\frac{\sin(k_{2}a)}{k_{2}}\tilde{P}(k_{2},\tau_{2};k_{1},\tau_{1})
    \label{ta2i}
\end{eqnarray}

Integrating over $k_2$, we again refer to appendix \ref{app:Ta} for details on the integration, we arrive to the expression
\begin{eqnarray}
    \left\langle T_{a}(t)^{2}\right\rangle =\frac{2}{\pi^{2}}\int_{0}^{t}d\tau_{2}\int_{0}^{\tau_{2}}d\tau_{1}\int_{-\infty}^{\infty}dk_{1}\frac{\sin(k_{1}a)}{k_{1}}e^{ik_{1}\left\langle x(\tau_{1})\right\rangle -\frac{k_{1}^{2}}{2}C(\tau_{1},\tau_{1})}I_{2}(a,k_{1},\tau_{1},\tau_{2}),
    \label{Ta2ex}
\end{eqnarray}
where $I_2(a,K_1,\tau_1,\tau_2)$ is defined as
$$
I_{2}(a,k_1,\tau_1,\tau_2)=\frac{\pi}{2}\left[\textrm{erf}\left(\frac{a-\left\langle x(\tau_{2})\right\rangle -ik_{1}C(\tau_{1},\tau_{2})}{\sqrt{2C(\tau_{2},\tau_{2})}}\right)+\textrm{erf}\left(\frac{a+\left\langle x(\tau_{2})\right\rangle +ik_{1}C(\tau_{1},\tau_{2})}{\sqrt{2C(\tau_{2},\tau_{2})}}\right)\right].
$$

To proceed further we consider that the random walk is isotropic, so that $\left\langle x(\tau_{i})\right\rangle =0$ with $i=1,2$. In this case, Eqs. \eqref{meantm} and \eqref{t2g} read
\begin{eqnarray}
    \left\langle T^{+}(t)\right\rangle &=&\frac{t}{2},\nonumber\\
    \left\langle T^{+}(t)^{2}\right\rangle &=&\frac{t^{2}}{2}-\frac{1}{\pi}\int_{0}^{t}d\tau_{2}\int_{0}^{\tau_{2}}d\tau_{1}\arctan\left(\sqrt{\frac{C(\tau_{2},\tau_{2})C(\tau_{1},\tau_{1})}{C(\tau_{1},\tau_{2})^{2}}-1}\right).
    \label{t1t2}
\end{eqnarray}
It is noteworthy that the results \eqref{t1t2} are exact general results for the first two moments of the half occupation time of a Gaussian isotropic process and can be explicitly found from the autocorrelation functions of the underlying random walk.
Analogously, for the occupation time in an interval, Eqs. \eqref{Ta1} and \eqref{Ta2ex} become
\begin{eqnarray}
    \left\langle T_{a}(t)\right\rangle =\int_{0}^{t}\textrm{erf}\left(\frac{a}{\sqrt{2C(\tau,\tau)}}\right)d\tau
    \label{Taei}
\end{eqnarray}
and
\begin{eqnarray}
    \left\langle T_{a}(t)^{2}\right\rangle &=&\frac{1}{\pi^{}}\int_{0}^{t}d\tau_{2}\int_{0}^{\tau_{2}}d\tau_{1}\int_{-\infty}^{\infty}dk\frac{\sin(ka)}{k}e^{-\frac{k_{}^{2}}{2}C(\tau_{1},\tau_{1})}\nonumber\\
    &\times &\left[\textrm{erf}\left(\frac{a-ik_{}C(\tau_{1},\tau_{2})}{\sqrt{2C(\tau_{2},\tau_{2})}}\right)+\textrm{erf}\left(\frac{a+ik_{}C(\tau_{1},\tau_{2})}{\sqrt{2C(\tau_{2},\tau_{2})}}\right)\right].
    \label{doserf}
\end{eqnarray}
In most cases the one-time autocorrelation (i.e., the mean square displacement of the random walk) $C(\tau,\tau)=\left\langle x(\tau)^2\right\rangle$ depends on time as $\left\langle x(\tau)^2\right\rangle \sim t^p$ in the long time limit. Depending on the value of $p$ we distinguish
subdiffusion ($0<p<1$) and superdiffusion ($p>1$). In any case, the argument of the error function in Eq. \eqref{Taei} defines the dimensionless quantity $a/\sqrt{2C(\tau,\tau)}$ (of order $t^{-p/2}$) that is very small in the long time limit. When the argument is small the error function can be approximated  as $\textrm{erf}(z)\simeq 2z/\sqrt{\pi}$. Then, Eq. \eqref{Taei} is
\begin{eqnarray}
    \left\langle T_{a}(t)\right\rangle \simeq a\sqrt{\frac{2}{\pi}}\int_{0}^{t}\frac{d\tau}{\sqrt{C(\tau,\tau)}}
    \label{Ta1ap}
\end{eqnarray}
in the long time limit. This result can be recovered making use of the infinite ergodic theory. To account for the occupation time in the interval $[-a,a]$ we take $U(x)=\mathds{1}_{[-a,a]}(x)$ into \eqref{mZ2} so that $\mathcal{A}=2a$ and \eqref{mZ2} reduces to \eqref{Ta1ap}.

Proceeding analogously with Eq. \eqref{t1t2}, we consider the term $a/\sqrt{2C(\tau_2,\tau_2)}$ much smaller than $kC(\tau_1,\tau_2)/\sqrt{2C(\tau_2,\tau_2)}$ in the long time limit. In this case we can approximate the error functions in Eq. \eqref{doserf} taking into account that $\textrm{erf}(\epsilon+iy)+\textrm{erf}(\epsilon-iy)\simeq 4\epsilon e^{y^{2}}/\sqrt{\pi}+O(\epsilon^{2})$. Hence,
$$
\left\langle T_{a}(t)^{2}\right\rangle \simeq4a\int_{0}^{t}\frac{d\tau_{2}}{\sqrt{2\pi C(\tau_{2},\tau_{2})}}\int_{0}^{\tau_{2}}\textrm{erf}\left(\frac{a}{\sqrt{2}}\sqrt{\frac{C(\tau_{2},\tau_{2})}{C(\tau_{1},\tau_{1})C(\tau_{2},\tau_{2})-C(\tau_{1},\tau_{2})^{2}}}\right)d\tau_{1}
$$
provided that $C(\tau_{1},\tau_{1})C(\tau_{2},\tau_{2})-C(\tau_{1},\tau_{2})^{2}>0$. Since the argument of the above error function is small in the long time limit we finally have
\begin{eqnarray}
    \left\langle T_{a}(t)^{2}\right\rangle \simeq\frac{4a^{2}}{\pi}\int_{0}^{t}d\tau_{2}\int_{0}^{\tau_{2}}\frac{d\tau_{1}}{\sqrt{C(\tau_{1},\tau_{1})C(\tau_{2},\tau_{2})-C(\tau_{1},\tau_{2})^{2}}}.
    \label{Ta2a}
\end{eqnarray}
 Likewise, the results \eqref{Ta1ap} and \eqref{Ta2a} correspond to the first and second moment of the occupation time in an interval for a Gaussian isotropic process in the long time limit, and like the half-occupation time, they depend solely on the autocorrelation function of the process.

\section{Applications}
Below, we explicit our results for specific examples of isotropic Gaussian processes of physical interest. In particular, we consider two isotropic Gaussian processes which generate anomalous diffusion paths. They are the scaled Brownian motion (SBM) and the fractional Brownian motion (fBM) \cite{Me14}. However, let us begin by considering the case of time-dependent diffusivity.

\subsection{Time-dependent diffusivity and SBM}
Consider that the position $x(t)$ of a random walker obeys the Langevin equation with time-dependent diffusivity 
\begin{equation}
    \frac{dx(t)}{dt}=\sqrt{2D(t)}\xi(t)
    \label{LE}
\end{equation}
where $\xi (t)$ is white Gaussian noise with zero mean and unit amplitude $\left\langle \xi(t)\xi(t')\right\rangle =\delta(t-t')$. Since $\xi(t)$ is Gaussian, so is $x(t)$ and its autocorrelation reads
\begin{eqnarray}
    C(t_{1},t_{2})=2\int_{0}^{t_{1}}\sqrt{D(\tau_{1})}d\tau_{1}\int_{0}^{t_{2}}\sqrt{D(\tau_{2})}\left\langle \xi(\tau_{1})\xi(\tau_{2})\right\rangle d\tau_{2}=2\int_{0}^{\min(t_{1},t_{2})}D(\tau)d\tau,
    \label{C12}
\end{eqnarray}
so that
$$
C(t,t)=2\int_{0}^{t}D(\tau)d\tau.
$$
The mean half occupation time is as in Eq. \eqref{t1t2} due to the isotropy of the random walk.  The mean square half occupation time follows from \eqref{t1t2}
using \eqref{C12}. Then,
\begin{eqnarray}
    \left\langle T^{+}(t)^{2}\right\rangle =\frac{t^{2}}{2}\left[1-\frac{2}{\pi}\int_{0}^{1}vdv\int_{0}^{1}\arctan\left(\sqrt{\frac{\int_{0}^{vt}D(\tau)d\tau}{\int_{0}^{uvt}D(\tau)d\tau}-1}\right)du\right],
    \label{tm22}
\end{eqnarray}
where we introduced the dimensionless variables $u=\tau_1/\tau_2$ and $v=\tau_2/t$. 
Analogously, we derive the first two moments for $T_a$ from \eqref{Ta1ap} and \eqref{Ta2a}. We thus find,
\begin{eqnarray}
    \left\langle T_{a}(t)\right\rangle \simeq\frac{at}{\sqrt{\pi}}\int_{0}^{1}\left[\int_{0}^{zt}D(\tau)d\tau\right]^{-1/2}dz
    \label{Ta2}
\end{eqnarray}
and
\begin{eqnarray}
    \left\langle T_{a}(t)^{2}\right\rangle \simeq\frac{2a^{2}t^{2}}{\pi}\int_{0}^{1}vdv\int_{0}^{1}du\left[\int_{0}^{uvt}D(\tau)d\tau\right]^{-1/2}\left[\int_{uvt}^{vt}D(\tau)d\tau\right]^{-1/2}.
    \label{Ta22}
\end{eqnarray}
The above expressions for the first two moments of $T^+$ and $T_a$ for a time-dependent diffusivity are general, depend on the expression for $D(t)$, and hold in the long time limit. 

A particularly interesting example of time-dependent diffusivity consists in assuming the dependence 
\begin{eqnarray}
    D(t) = \alpha K t^{\alpha-1}
    \label{Dsbm}
\end{eqnarray}
with $\alpha \in (0,2)$. This case corresponds to the so called Scaled Brownian Motion (SBM). In this case, the two-time autocorrelation function and the mean square displacement (MSD) are, respectively
\begin{eqnarray}
C(t_{1},t_{2})=2K\left[\min(t_{1},t_{2})\right]^{\alpha},\quad\left\langle x(t)^{2}\right\rangle =2Kt^{\alpha},
    \label{cmsd}
\end{eqnarray}
which can be derived integrating \eqref{LE} with \eqref{Dsbm}. SBM was used to describe fluorescence recovery after photobleaching in various settings \cite{Sa01}
as well as anomalous diffusion in various biophysical contexts \cite{Sz06,Wu08,Me14}. In other branches of
physics SBM was used to study the scaled voter model \cite{Ka23} and to model turbulent flows observed by Richardson \cite{Ri26}. Moreover, the diffusion of particles in granular gases with
relative speed dependent restitution coefficients follow SBM \cite{Bo15}. The ergodic properties of SBM have also been studied considering the square displacement as the observable \cite{Sa15}.  SBM has short memory and its increments are not stationary, which is particularly prominent under confinement \cite{Je14,ChMe15}.

\subsubsection{Half occupation time}

\begin{figure}[!htbp]
    \centering
    \includegraphics[width=0.5\linewidth]{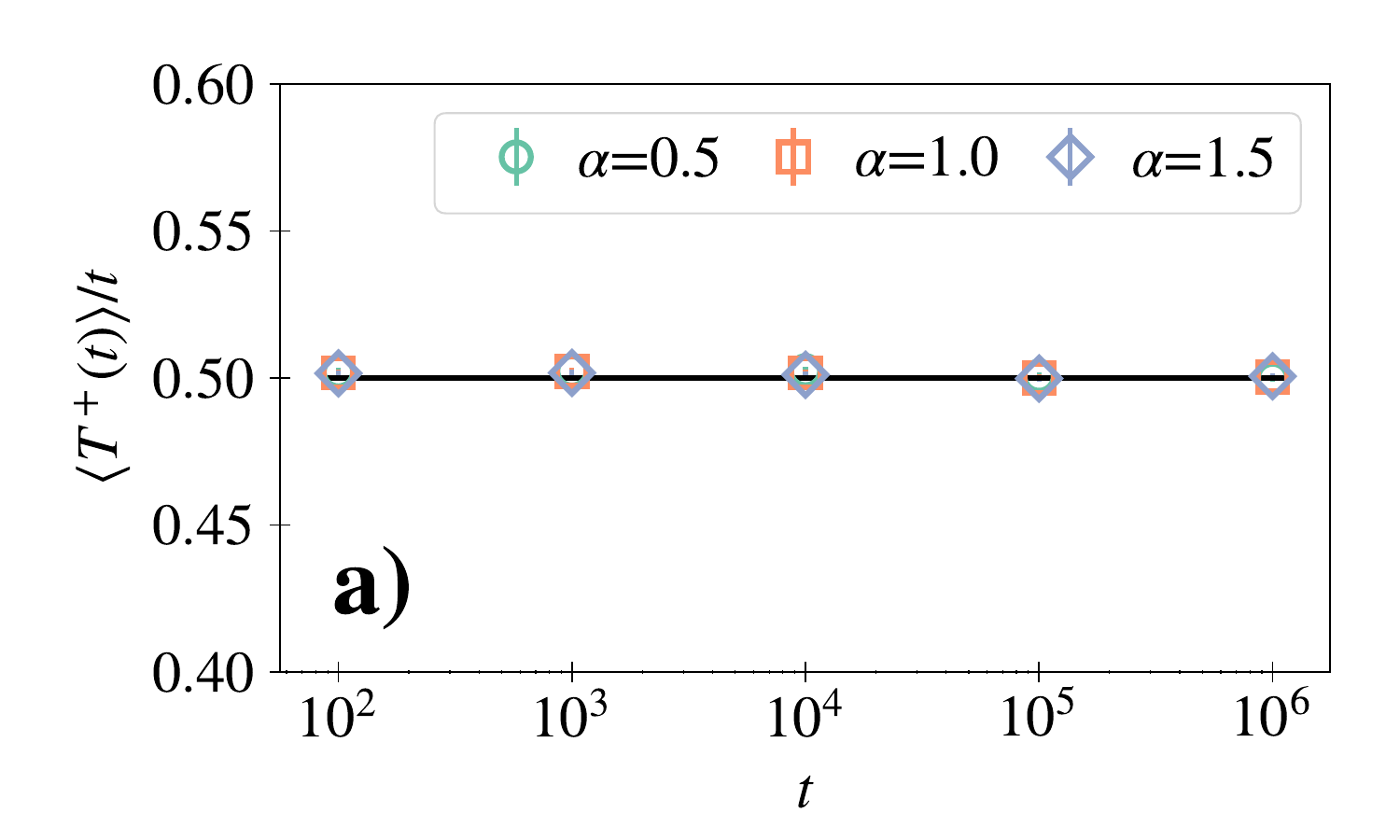}%
    \includegraphics[width=0.5\linewidth]{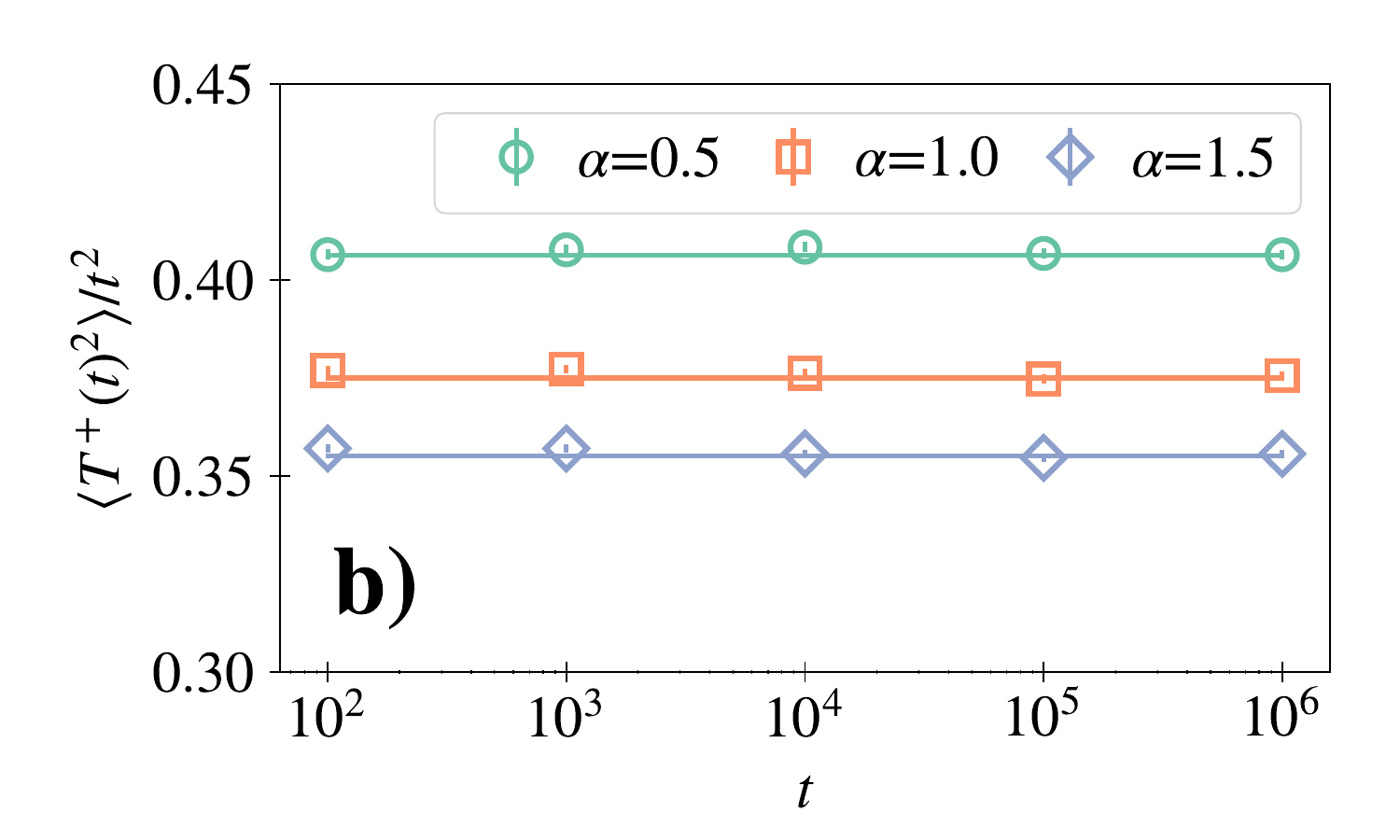}
    \caption{First -a)- and second -b)- moments of the half occupation time over $t$ and $t^2$ respectively as a function of $t$ for the SBM. The symbols are obtained from numerical simulations for $\alpha=0.5,1.0,$ and $1.5$. The solid lines are given by Eq. \eqref{t1t2} in panel a), and by Eq. \eqref{eq:SBM_TPl2} in panel b). All simulations are performed with parameters: $x_0=0$, $K=1$, time discretization of $dt=0.1$, and $N=10^5$ trajectories.}
    \label{fig:SBM_Tpls}
\end{figure}

We consider here the half occupation time $T^+(t)$ for the SBM. From \eqref{t1t2} and \eqref{cmsd} 
\begin{eqnarray}
\left\langle T^{+}(t)^{2}\right\rangle &=&\frac{t^{2}}{2}-\frac{1}{\pi}\int_{0}^{t}d\tau_{2}\int_{0}^{\tau_{2}}d\tau_{1}\arctan\left(\sqrt{\frac{\tau_{2}^{\alpha}}{\tau_{1}^{\alpha}}-1}\right)\nonumber\\
&=&\frac{t^{2}}{2}\left[1-\frac{1}{\pi}\int_{0}^{1}\arctan\left(\sqrt{\frac{1}{u^{\alpha}}-1}\right)du\right].
\label{tm2}
\end{eqnarray}
The integral in the above expression can be solved by introducing the new variable $y=(u^{-\alpha}-1)^{1/2}$
$$
\int_{0}^{1}\arctan\left(\sqrt{\frac{1}{u^{\alpha}}-1}\right)du=\frac{2}{\alpha}\int_{0}^{\infty}\frac{y\arctan(y)}{(1+y^{2})^{1+\frac{1}{\alpha}}}dy=\frac{\sqrt{\pi}}{2}\frac{\Gamma\left(\frac{1}{2}+\frac{1}{\alpha}\right)}{\Gamma\left(1+\frac{1}{\alpha}\right)}
$$
so that Eq. \eqref{tm2} reduces to
\begin{eqnarray}
    \left\langle T^{+}(t)^{2}\right\rangle =\frac{t^{2}}{2}\left[1-\frac{1}{2\sqrt{\pi}}\frac{\Gamma\left(\frac{1}{2}+\frac{1}{\alpha}\right)}{\Gamma\left(1+\frac{1}{\alpha}\right)}\right].
    \label{eq:SBM_TPl2}
\end{eqnarray}
Note that the case for constant diffusivity is recovered when $\alpha=1$. Indeed, setting $\alpha=1$ in the above expression we obtain $\left\langle T^{+}(t)^{2}\right\rangle =3t^2/8$ which is the result corresponding to the standard Brownian motion \cite{BaFlMe23,MeFlPa25}. Using \eqref{EB2} the EB parameter is given by
\begin{eqnarray}
\textrm{EB}_{+}=\frac{\left\langle T^{+}(t)^{2}\right\rangle }{\left\langle T^{+}(t)\right\rangle ^{2}}-1=1-\frac{1}{\sqrt{\pi}}\frac{\Gamma\left(\frac{1}{2}+\frac{1}{\alpha}\right)}{\Gamma\left(1+\frac{1}{\alpha}\right)}
\label{eq:SBM_EBp}
\end{eqnarray}
which is monotonically decreasing with $\alpha$. Note that for $\alpha=1$ one has $\textrm{EB}_{+}=1/2$ which is the result corresponding to the standard Brownian motion \cite{MeFlPa25}. 

In Figure \ref{fig:SBM_Tpls} we compare numerical simulations of SBM, where we have generated a set of trajectories using the Milstein algorithm to solve Eq. \eqref{LE} with Eq. \eqref{Dsbm}, with the results of the first and second moments of the half occupation time as a function of time, for three different values of the exponent $\alpha$. In all cases, there is a good agreement between the simulations and the analytical results.  In panel a) of Figure \ref{fig:SBM_EB} we present the EB parameter for this observable, as a function of the exponent $\alpha$. Again, we find a good agreement between Eq. \eqref{EB3} and the numerical data.

\subsubsection{Occupation time in an interval}

\begin{figure}[!htbp]
    \centering
    \includegraphics[width=0.5\linewidth]{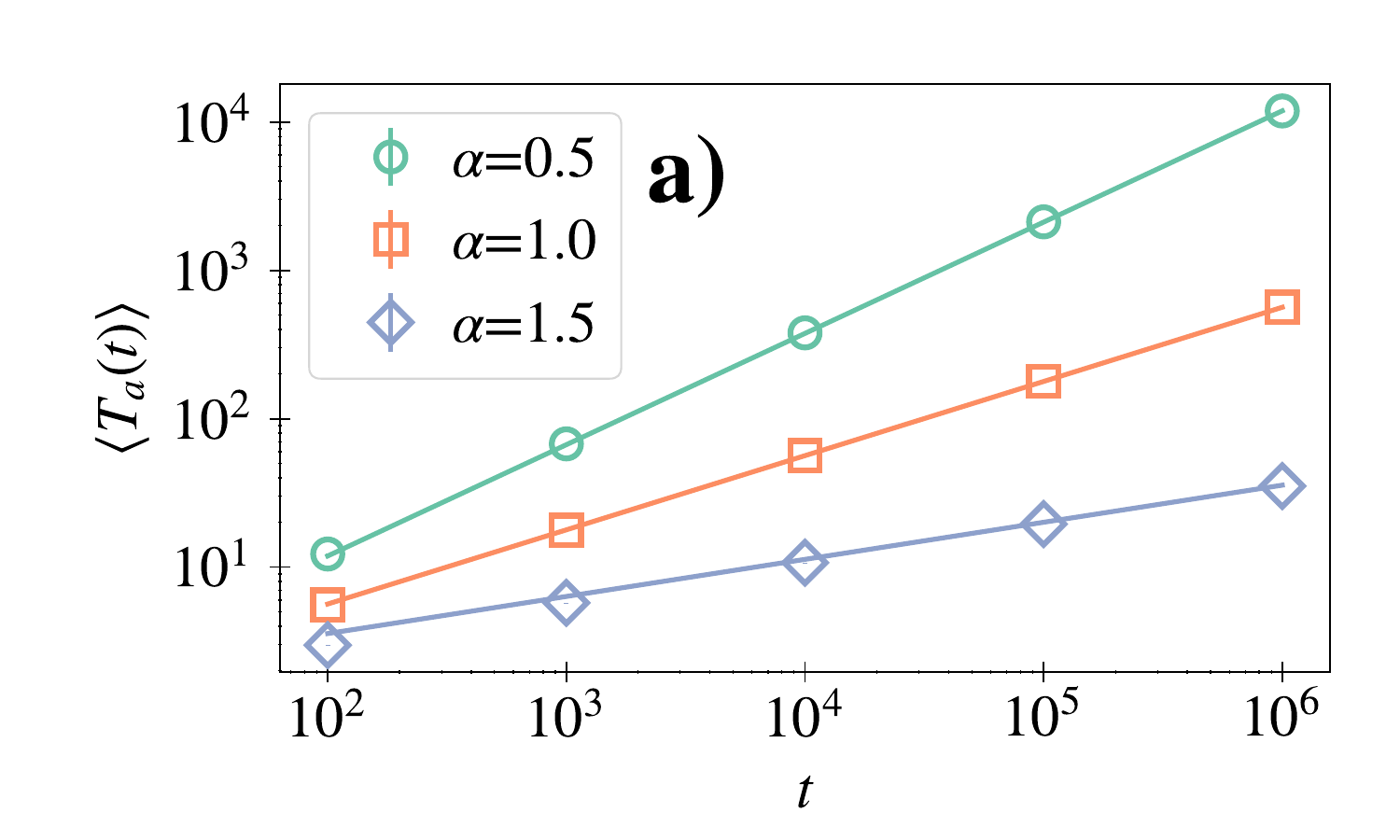}%
    \includegraphics[width=0.5\linewidth]{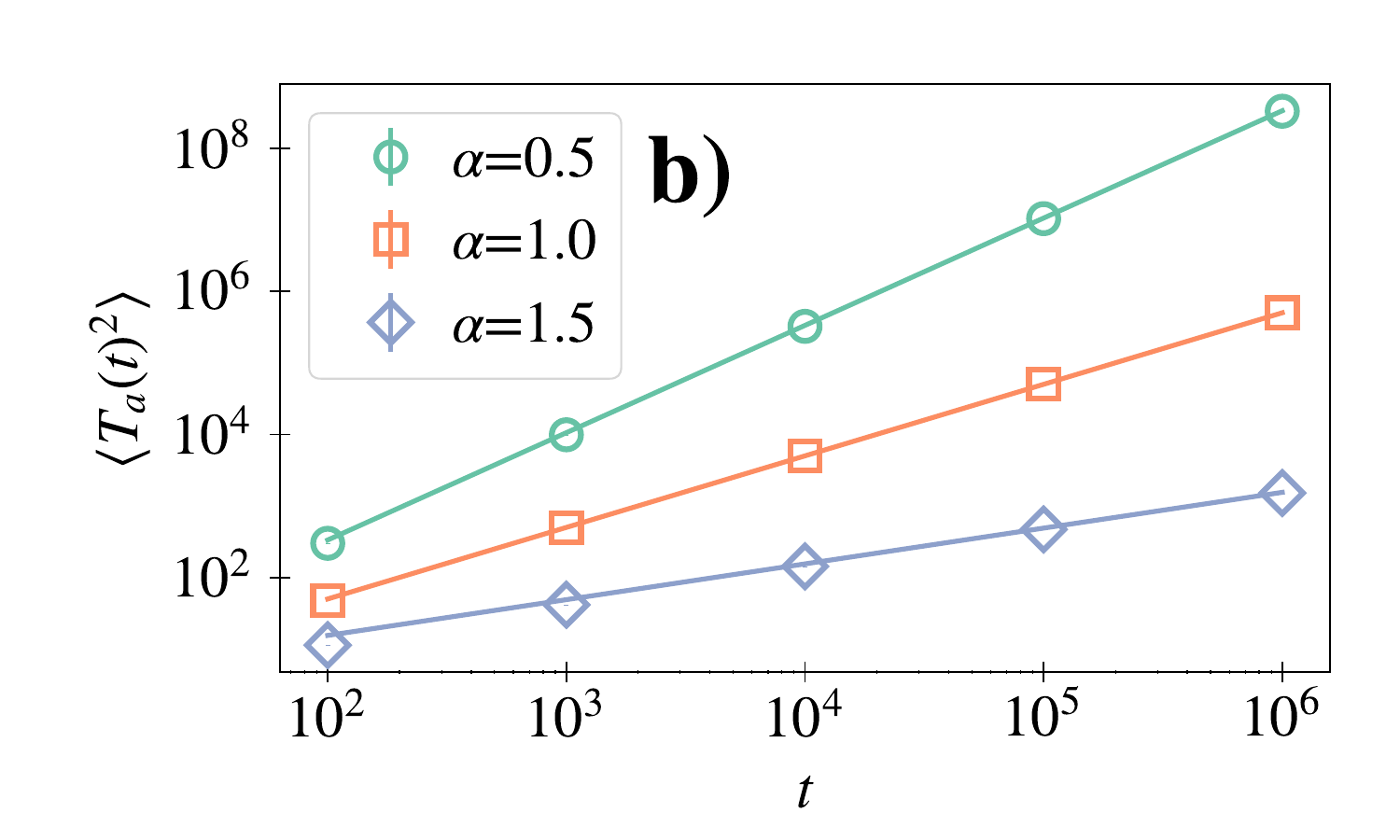}
    \caption{First -a)- and second -b)- moments of the occupation time in an interval as a function of $t$ for the SBM. The symbols are obtained from numerical simulations for $\alpha=0.5,1.0,$ and $1.5$. The solid lines are given by Eq. \eqref{talt} in panel a), and by Eq. \eqref{eq:SBM_Ta2} in panel b). All simulations are performed with parameters: $a=0.5$, $x_0=0$, $K=1$, time discretization of $dt=0.1$, and $N=10^5$ trajectories.}
    \label{fig:SBM_Ta}
\end{figure}

From \eqref{Ta2} we can obtain the mean occupation time in an interval for a SBM. Using \eqref{cmsd} and \eqref{Ta2} we readily find
\begin{eqnarray}
\left\langle T_{a}(t)\right\rangle \simeq\frac{2at^{1-\frac{\alpha}{2}}}{(2-\alpha)\sqrt{\pi K}}\quad\text{as}\quad t\to\infty.
\label{talt}
\end{eqnarray}
For $\alpha=1$ we recover the result $\left\langle T_{a}(t)\right\rangle \simeq 2a\sqrt{t}/\sqrt{\pi D}$ corresponding to the standard Brownian motion \cite{BaFlMe23,MeFlPa25}. Note that considering  $U(x)=\mathds{1}_{[-a,a]}(x)$ and \eqref{cmsd} into \eqref{mZ2}
we recover \eqref{talt}.

The mean square occupation time in an interval follows from \eqref{Ta22} and \eqref{cmsd}. We find
\begin{eqnarray}
    \left\langle T_{a}(t)^{2}\right\rangle \simeq\frac{2a^{2}}{K\sqrt{\pi}}\frac{\Gamma\left(\frac{1}{\alpha}-\frac{1}{2}\right)}{\alpha(2-\alpha)\Gamma(1/\alpha)}t^{2-\alpha}\quad\text{as}\quad t\to \infty.
    \label{eq:SBM_Ta2}
\end{eqnarray}
Note that for $\alpha=1$ we get $\left\langle T_{a}(t)^{2}\right\rangle \simeq 2a^2t/D$ which corresponds to the mean square occupation time for the standard Brownian motion \cite{BaFlMe23,MeFlPa25}.
The EB parameter is from \eqref{EB2} given by
\begin{eqnarray}
\textrm{EB}_{a}=\frac{\left\langle T_{a}(t)^{2}\right\rangle }{\left\langle T_{a}(t)\right\rangle ^{2}}-1=\frac{\sqrt{\pi}\Gamma\left(\frac{1}{\alpha}-\frac{1}{2}\right)(2-\alpha)}{2\alpha\Gamma(1/\alpha)}-1
\label{eq:SBM_EBa}
\end{eqnarray}
which is monotonically decreasing with $\alpha$. For $\alpha=1$ we have $\text{EB}_a=\pi /2-1$, which corresponds to the result for the standard Brownian motion \cite{BaFlMe23}.  
From previous results, it is known that the PDF of $T_a(t)$ follows the Mittag-Leffler (ML) distribution of order $\beta$, where $\beta$ is obtained from the long time limit of the First Passage Time PDF $\sim t^{-1-\beta}$, if the return times to the interval are a renewal process \cite{Darling_1957,GoLu01,LeBa19,He08,Ki22}.
From these studies it follows that the ergodicity breaking parameter  corresponding to the ML distribution, EB$^*_a$, is
\begin{eqnarray}
\textrm{EB}^{*}_{a}=\frac{2\Gamma^2(1+\beta)}{\Gamma(1+2\beta)}-1, 
\label{eq:EB_ML}
\end{eqnarray}
which, for the SBM, $\beta=\alpha/2$ \cite{Safd2015}. Comparing Eq. \eqref{eq:EB_ML} to our result Eq. \eqref{eq:SBM_EBa} we see that the two expressions do not yield the same predictions for $\mathrm{EB}_a$. Therefore, the return times of the SBM are not, in general, a renewal process, unless $\alpha=1$.

\begin{figure}[!htbp]
    \centering
    \includegraphics[width=0.5\linewidth]{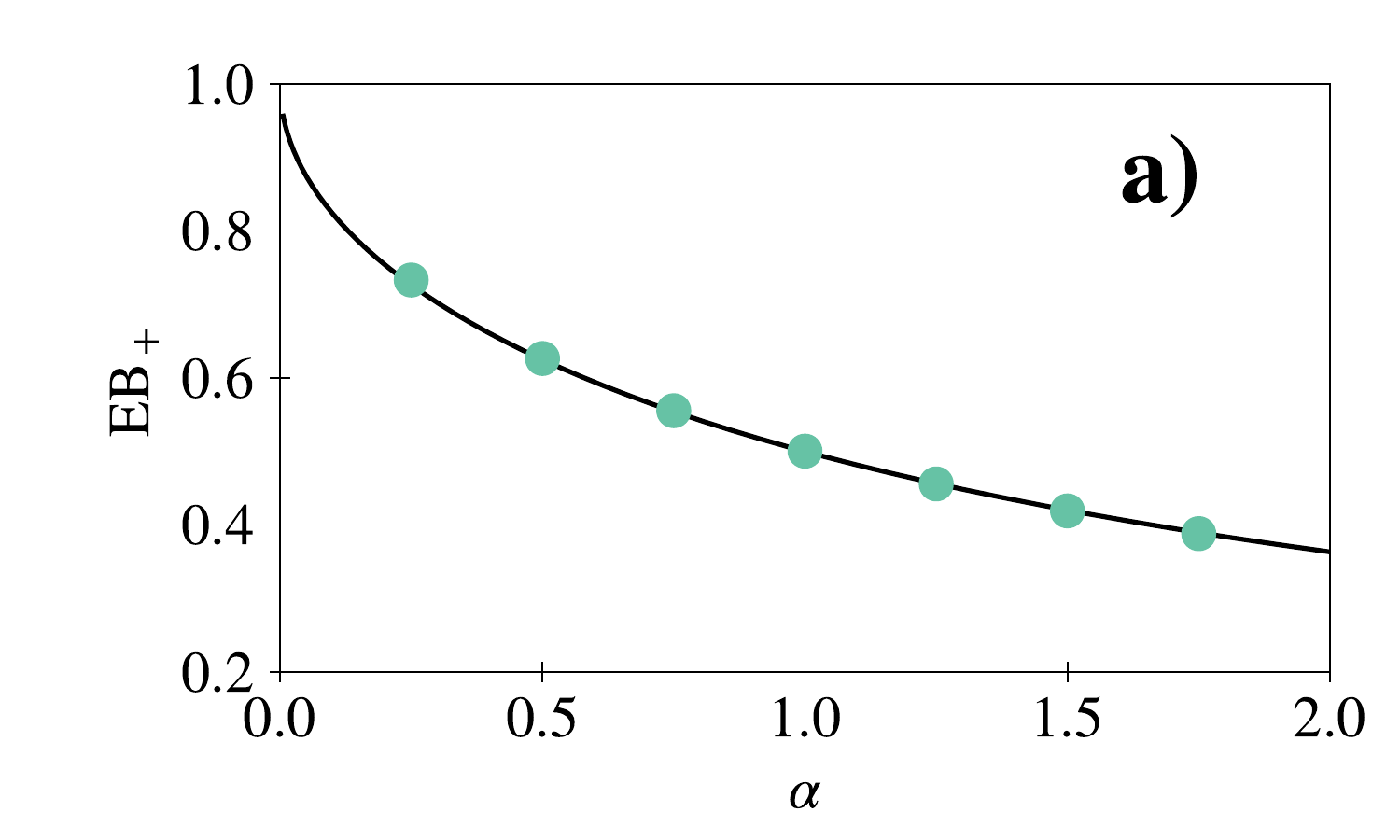}%
    \includegraphics[width=0.5\linewidth]{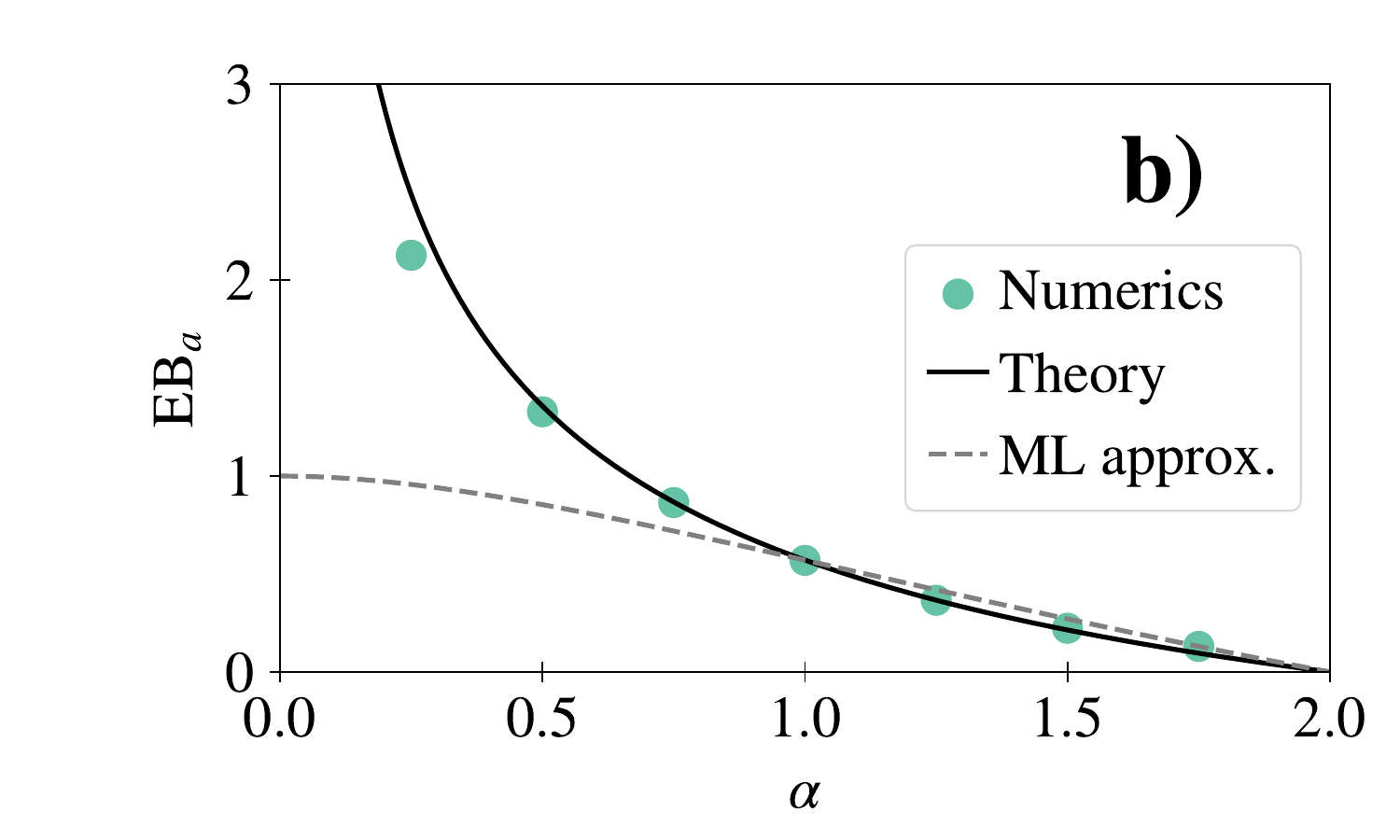}
    \caption{Ergodicity breaking parameter EB as function of the exponent $\alpha$ for the half occupation time -a)- and the occupation time in an interval -b)-. The symbols are computed with numerical simulations, and the solid lines correspond to Eq. \eqref{eq:SBM_EBp} in panel a) and to Eq. \eqref{eq:SBM_EBa} in panel b). The dashed line in panel b) corresponds to Eq. \eqref{eq:EB_ML} with $\beta=\alpha/2$. All simulations are performed with parameters: $a=0.5$, $x_0=0$, $K=1$, time discretization of $dt=0.1$, and $N=10^5$ trajectories.}
    \label{fig:SBM_EB}
\end{figure}

We compare the results of this section with numerical simulations in Figure \ref{fig:SBM_Ta} and in panel b) of Figure \ref{fig:SBM_EB}. In the first, we plot the first and second moments of the occupation time in the interval $[-0.5,0.5]$ as a function of time, for distinct values of the exponent $\alpha$. We find a good agreement between the analytic results and the numerical data. We also observe that the total trajectory time required for the simulation and the data to converge is smaller for $\alpha \to 1$, and increases for $\alpha \to 0$ and $\alpha \to 2$. This can also be seen in Figure \ref{fig:SBM_EB}b) where the agreement is better for $\alpha\to1$ for a given time $t$. We have also included in panel b) the prediction from Eq. \eqref{eq:EB_ML}, and we can see that for values of $1<\alpha<2$ the results provided by the ML distribution are quite similar to the exact theoretical prediction given by Eq. \eqref{eq:SBM_EBa}, being equal for $\alpha=1$, as expected. However, the approximation breaks for $0<\alpha<1$, and the numerical simulations, in agreement with \eqref{eq:SBM_EBa}, confirm the deviation as $\alpha\to0$.

\textcolor{black}{In Figure \ref{fig:SBM_EB}, both $\mathrm{EB}_+$ and $\mathrm{EB}_a$ decrease monotonically with $\alpha$. Because $\mathrm{EB}$ measures trajectory-to-trajectory variability in the observable $\mathcal{O}$, this implies greater trajectory dispersion at low $\alpha$. For $\alpha\in(0,1)$, the decreasing diffusion coefficient $D(t)$ makes escaping from a given region progressively harder, enhancing trajectory differences with respect to $T^+$ and $T_a$. Conversely, for $\alpha\in(1,2)$, $D(t)$ increases with time, reducing trajectory persistence and thus lowering EB.}

\subsection{Fractional Brownian motion}
The fractional Brownian motion (fBM) is a self-similar Gaussian process with zero mean and autocorrelation
\begin{eqnarray}
C(t_1,t_2)=D_{H}\left(t_{1}^{2H}+t_{2}^{2H}-|t_{1}-t_{2}|^{2H}\right)
\label{C}
\end{eqnarray}
with $H\in (0,1)$, so that $\left\langle x(t)^{2}\right\rangle=2D_Ht^{2H}$. 
It was first introduced by Kolmogorov \cite{Ko40} and later studied by Mandelbrot and van Ness \cite{Ma68}. It has been observed to generate the subdiffusion of various tracers in complex environments both in vivo and in vitro \cite{Ma09,Je11,Je12,Je13,Ta13}, but also for completely different stochastic processes such
as electronic network traffic \cite{Mi02} or financial time series \cite{Co98,Bi08}. In the superdiffusive
regime ($1/2<H<1$), positive increment correlations and single trajectory powerspectra
consistent with fBM were observed for the actively driven motion of endogenous
granules inside amoeba cells as well as for the motion of the amoeba themselves \cite{Re15,Kr19}. In addition it has been extensively studied in mathematical
literature \cite{Bi08,Qi08}. The statistical properties of fBM have also been studied. They include the ergodicity breaking, using the time-average square displacement, \cite{DeBa09,Je10}, the generalization of the arcsine laws \cite{Sa18} and the PDF in confined geometries \cite{Je10,Wi11,Vo21}. 
More recently, the ergodic properties for the occupation time of fBM has been analyzed. It has been shown that the EB parameter can be obtained from the ML distribution, (Eq. \eqref{eq:EB_ML} with order $\beta=1-H$), solely near the Brownian limit $H=1/2$ \cite{Ki22}. So that an exact expression for the EB valid for any $H$ was lacking. Below, we find exact analytic expressions for the EB of the occupation times of fBM. 
 
\subsubsection{Half occupation time}

\begin{figure}[!htbp]
    \centering
    \includegraphics[width=0.5\linewidth]{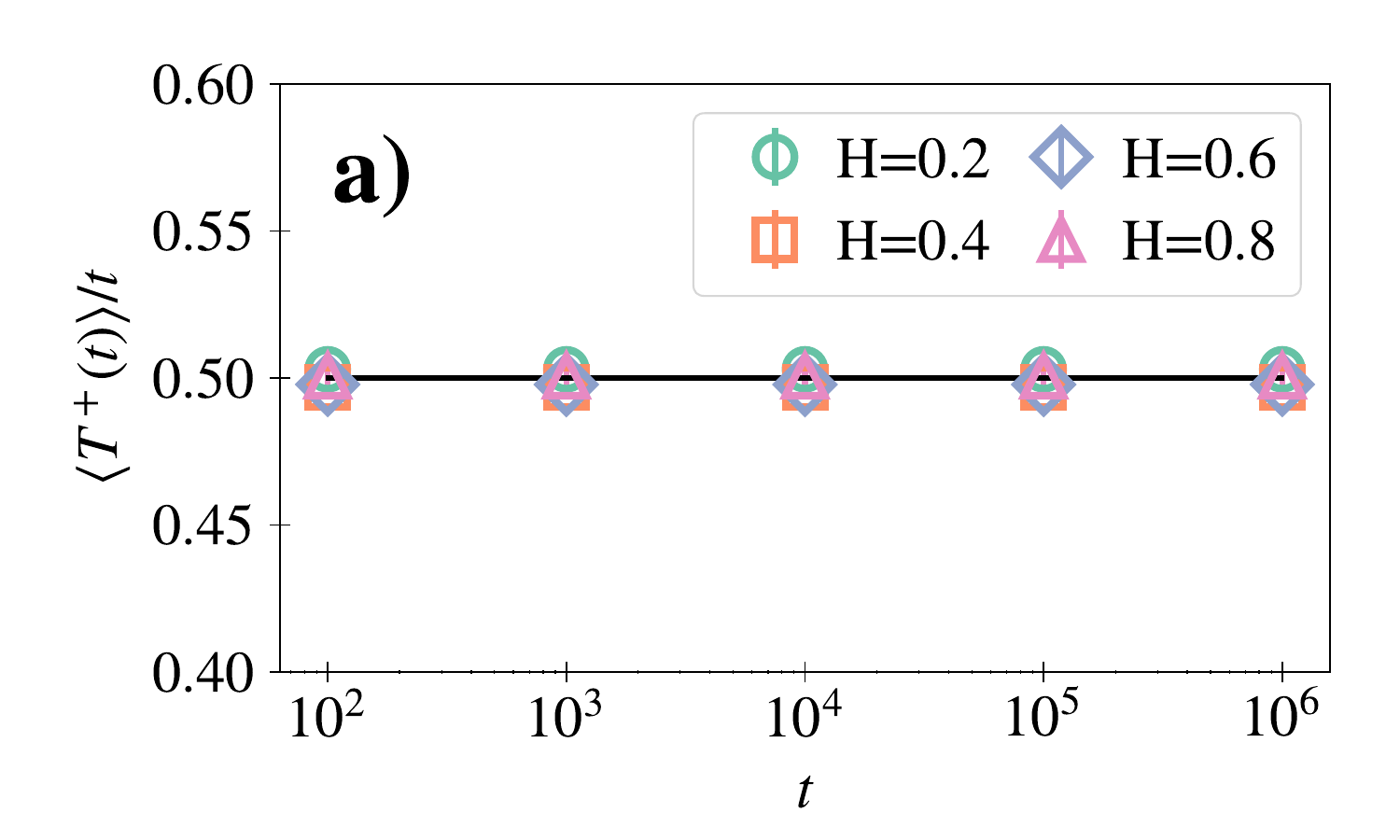}%
    \includegraphics[width=0.5\linewidth]{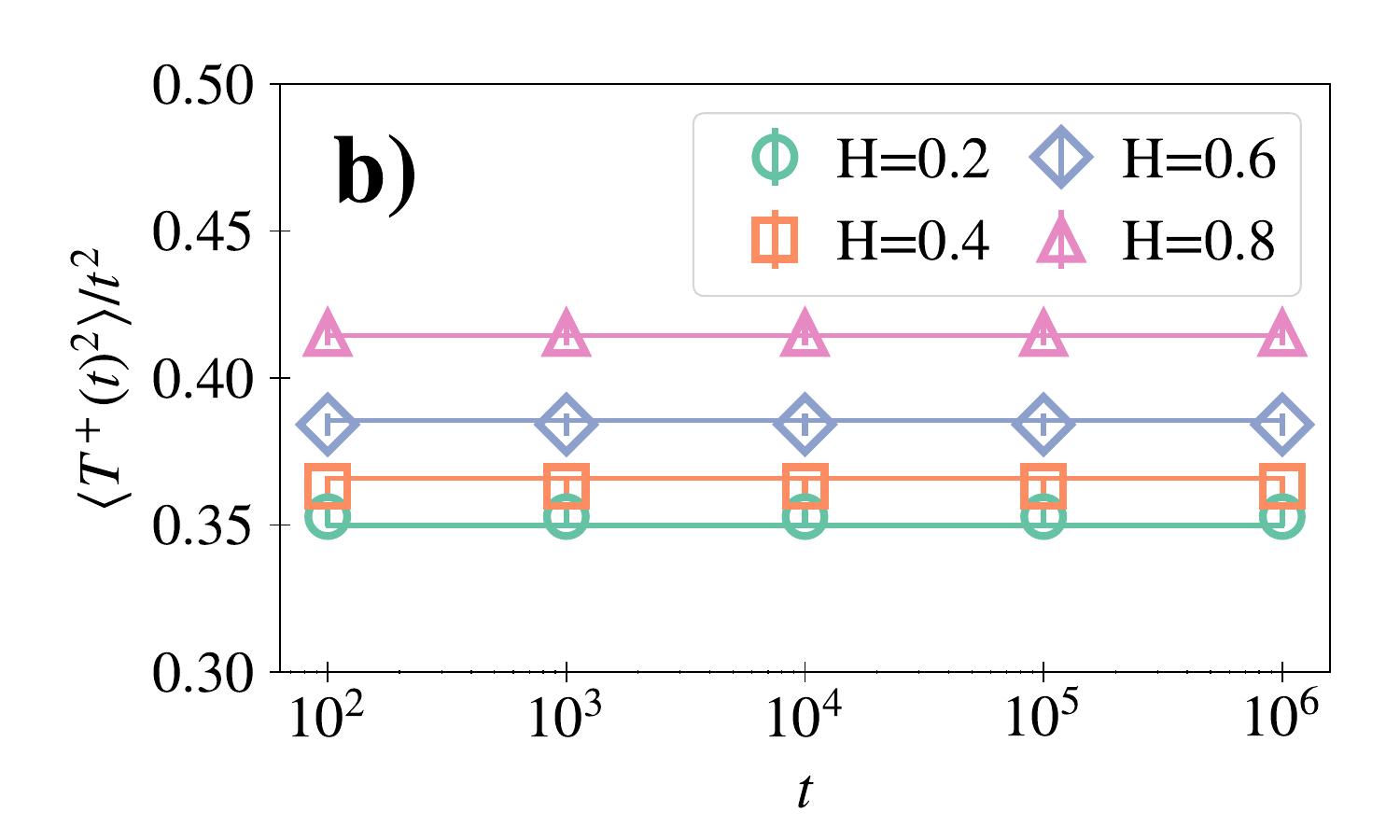}
    \caption{First -a)- and second -b)- moments of the half occupation time over $t$ and $t^2$ respectively as a function of $t$ for the fBM. The symbols are obtained from numerical simulations for $H=0.2,0.4,0.6,$ and $0.8$. The solid lines are given by Eq. \eqref{t1t2} in panel a), and Eq. \eqref{eq:fbmTp2} in panel b), where the integral has been computed numerically. All simulations are performed with parameters: $x_0=0$, $D_H=1/2$, number of time points $M=2^{23}$, and $N=10^4$ trajectories.}
    \label{fig:fBM_Tpls}
\end{figure}

The mean half occupation time is again given in Eq. \eqref{t1t2} while the mean square half occupation time follows from \eqref{t1t2} and \eqref{C}.  Finally, we find
\begin{eqnarray}
\left\langle T^{+}(t)^{2}\right\rangle =\frac{t^{2}}{2}\left(1-\frac{\mathcal{A}_{H}}{\pi}\right)
\label{eq:fbmTp2}
\end{eqnarray}
where
$$
\mathcal{A}_{H}=\int_{0}^{1}\arctan\left(\frac{\sqrt{4u^{2H}-\left[1+u^{2H}-(1-u)^{2H}\right]^{2}}}{1+u^{2H}-(1-u)^{2H}}\right)du.
$$
The ergodicity breaking parameter $\textrm{EB}_+$ is computed from \eqref{EB2} to get
\begin{eqnarray}
\textrm{EB}_{+}=1-\frac{2}{\pi}\mathcal{A}_{H}
    \label{EB3}
\end{eqnarray}
Note that on setting $H=1/2$ into \eqref{EB3} we recover $\textrm{EB}_{+}=1/2$. 

In Figure \ref{fig:fBM_Tpls} we compare numerical simulations of fBM, where we have obtained the set of trajectories using the circulant embedding method (see for example \cite{CBM-fBM}), with the results of the first and second moments of the half occupation time as a function of time, for distinct values of the exponent $H$. In panel a) of Figure \ref{fig:fBM_EB} we present the EB for this observable, as a function of the exponent $H$. We find a good agreement between Eq. \eqref{EB3} and the numerical data.

\subsubsection{Occupation time in an interval}

\begin{figure}[!htbp]
    \centering
    \includegraphics[width=0.5\linewidth]{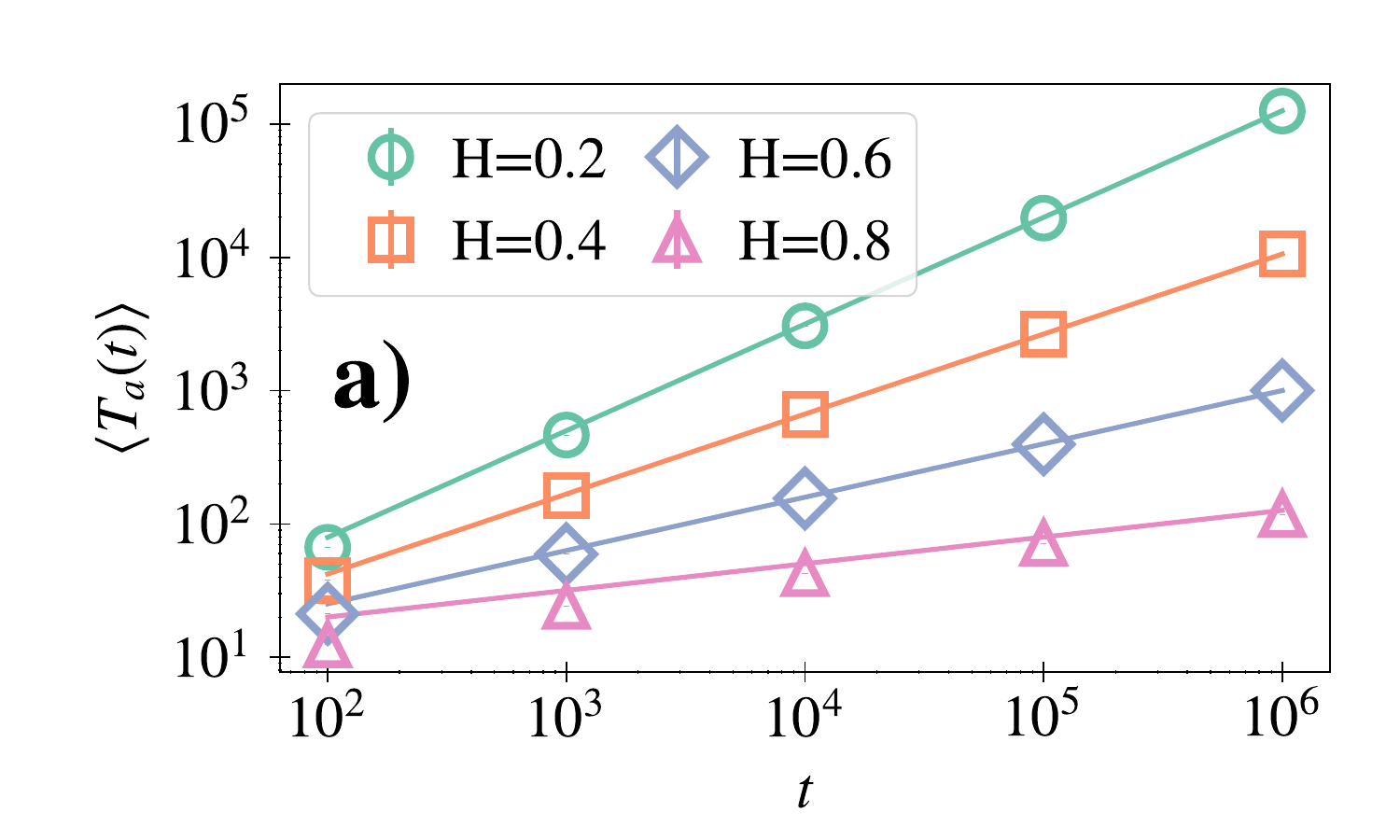}%
    \includegraphics[width=0.5\linewidth]{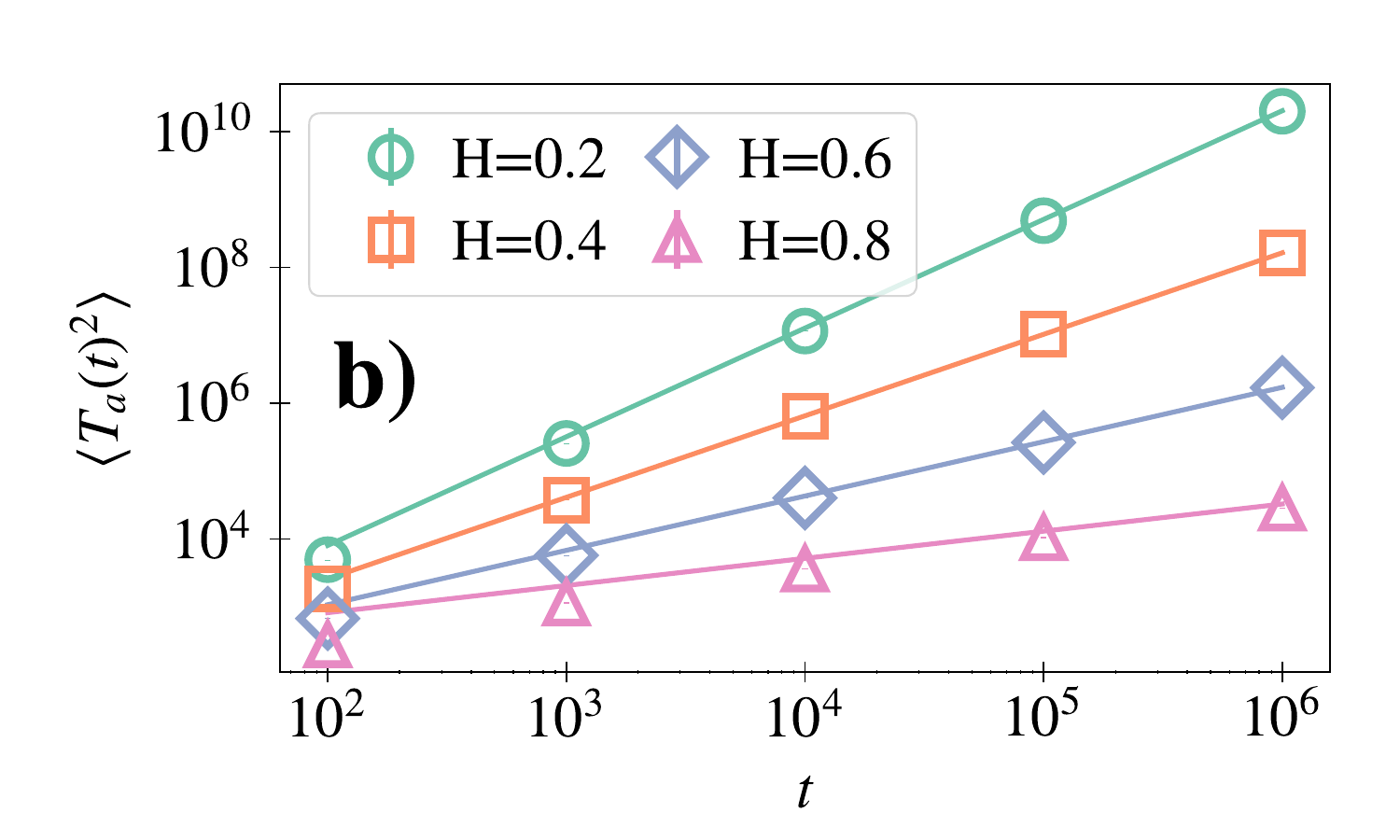}
    \caption{First -a)- and second -b)- moments of the occupation time in an interval as a function of $t$ for the fBM. The symbols are obtained from numerical simulations for $H=0.2,0.4,0.6,$ and $0.8$. The solid lines are given by Eq. \eqref{talt2} in panel a), and by Eq. \eqref{ta24} in panel b), where the integral has been computed numerically. All simulations are performed with parameters: $a=2$, $x_0=0$, $D_H=1/2$, number of time points $M=2^{23}$, and $N=10^4$ trajectories.}
    \label{fig:fBM_Ta}
\end{figure}

Although the stochastic dynamics of the SBM is not equivalent to that of the fBm, the mean occupation time is as in Eq. \eqref{talt} but replacing $\alpha$ and $K$ by $2H$ and $D_H$ respectively. Thus,
\begin{eqnarray}
\left\langle T_{a}(t)\right\rangle \simeq\frac{at^{1-H}}{(1-H)\sqrt{\pi D_{H}}}\quad\text{as}\quad t\to\infty .
    \label{talt2}
\end{eqnarray}
Note that considering $U(x)=\mathds{1}_{[-a,a]}(x)$ and \eqref{C} into \eqref{mZ2}
we recover \eqref{talt2}.

\begin{figure}[!htbp]
    \centering
    \includegraphics[width=0.5\linewidth]{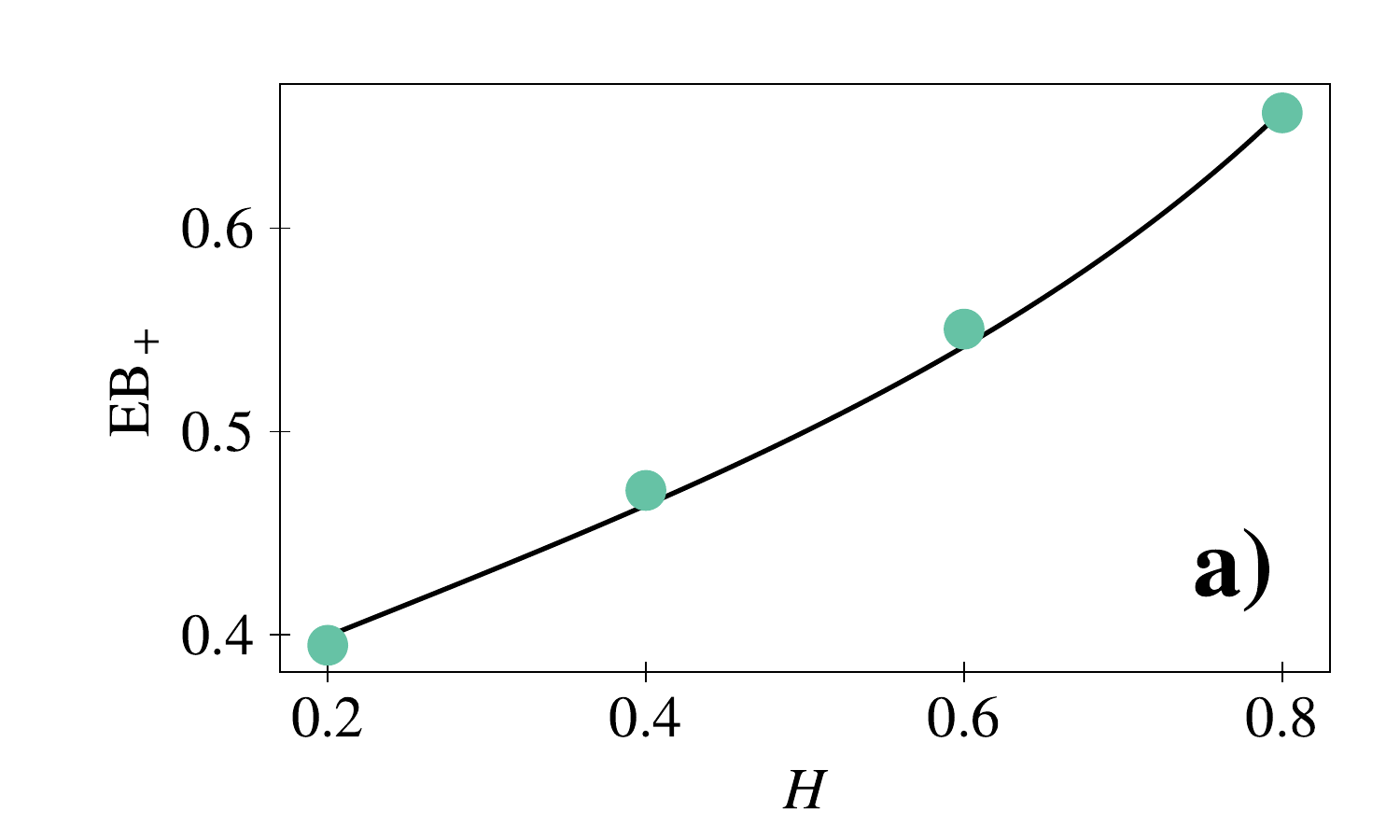}%
    \includegraphics[width=0.5\linewidth]{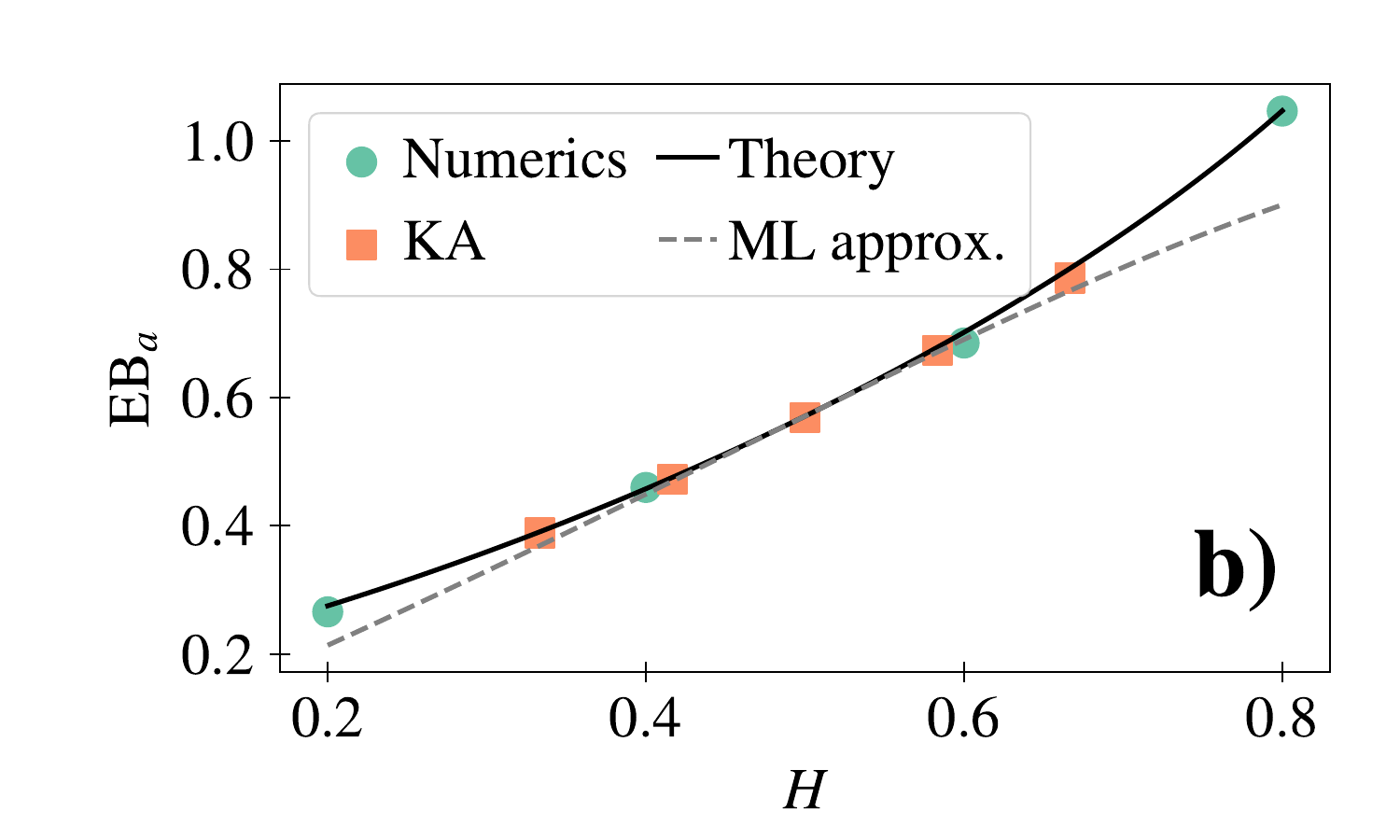}
    \caption{Ergodicity breaking parameter EB as function of the exponent $H$ for the half occupation time -a)- and the occupation time in an interval -b)-. The round green symbols are computed with numerical simulations, and the solid lines correspond to Eq. \eqref{EB3} in panel a) and to Eq. \eqref{eq:fBM_EB_ta} in panel b). In both cases, the integrals have been numerically computed. The dashed line in panel b) corresponds to the ML approximation \eqref{eq:EB_ML} and the orange squares to data from \cite{Ki22}. All our simulations are performed with parameters: $a=2$, $x_0=0$, $D_H=1/2$, number of time points $M=2^{23}$, final time $t=10^6$, and $N=10^4$ trajectories.  }
    \label{fig:fBM_EB}
\end{figure}

The mean square occupation time in the long time limit can be obtained after introducing \eqref{C} in \eqref{Ta2a} and considering the changes of variables $\tau_2=st$ ($s\in[0,1]$) and $\tau_1=u\tau_2$ ($u\in[0,1]$). We finally get
\begin{eqnarray}
\left\langle T_{a}(t)^{2}\right\rangle \simeq\frac{2a^{2}\mathcal{B}_{H}}{\pi D_{H}(1-H)}t^{2-2H}
    \label{ta24}
\end{eqnarray}
with
$$
\mathcal{B}_{H}=\int_{0}^{1}\frac{du}{\sqrt{4u^{2H}-\left[1+u^{2H}-(1-u)^{2H}\right]^{2}}}.
$$
The EB$_a$ follows easily from \eqref{EB2} together with \eqref{talt2} and \eqref{ta24}
\begin{eqnarray}
\textrm{EB}_{a}=2(1-H)\mathcal{B}_{H}-1.
\label{eq:fBM_EB_ta}
\end{eqnarray}
Note that for $H=1/2$ we find $\textrm{EB}_a=\pi/2-1$ as it should for the Brownian motion \cite{BaFlMe23}. 
In Figure \ref{fig:fBM_Ta} we compare numerical simulations of fBM with the results of the first and second moments of the occupation time in the interval $[-2,2]$ as a function of time, for distinct values of the exponent $H$. We find a good agreement between the analytic results and the numerical data. We also see that the convergence to the prediction is slower for higher values of $H$. We plot the EB$_a$ parameter in panel b) of Figure \ref{fig:fBM_EB}. We find a good agreement between Eq. \eqref{eq:fBM_EB_ta} and the numerical data. We observe that, as expected from the results in \cite{Ki22} (which we include in the plot as orange squares with the label KA), the EB$_a$ computed from the ML distribution is valid solely in the region near $H=1/2$, where Eq. \eqref{eq:fBM_EB_ta} and \eqref{eq:EB_ML} provide similar results. For values of $H\to0$ and $H\to1$ we find that \eqref{eq:EB_ML} is no longer valid. However, our result \eqref{eq:fBM_EB_ta} is exact and holds for the whole range of values of $H$.

\textcolor{black}{Figure \ref{fig:fBM_EB} shows that EB increases with $H$ for the fBM, which is consistent with the results for the SBM previously discussed, where greater persistence led to larger EB values. Since for $H > 1/2$ the fBM has positively correlated increments, the process is more persistent in this regime, leading to correspondingly higher values of EB for both $T_a$ and $T^+$.}

\subsection{Scaling form}

\begin{figure}[!htbp]
    \centering
    \includegraphics[width=0.5\linewidth]{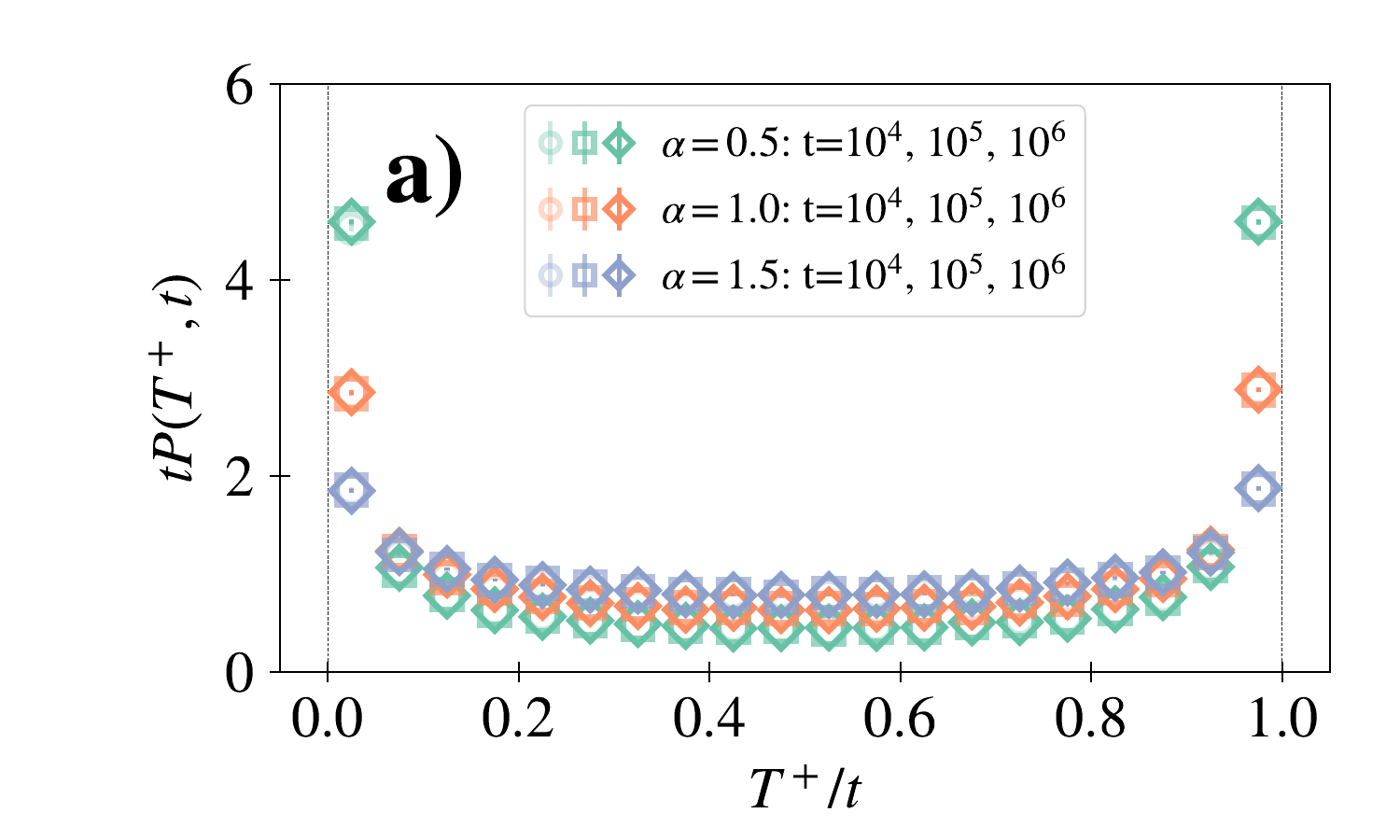}%
    \includegraphics[width=0.5\linewidth]{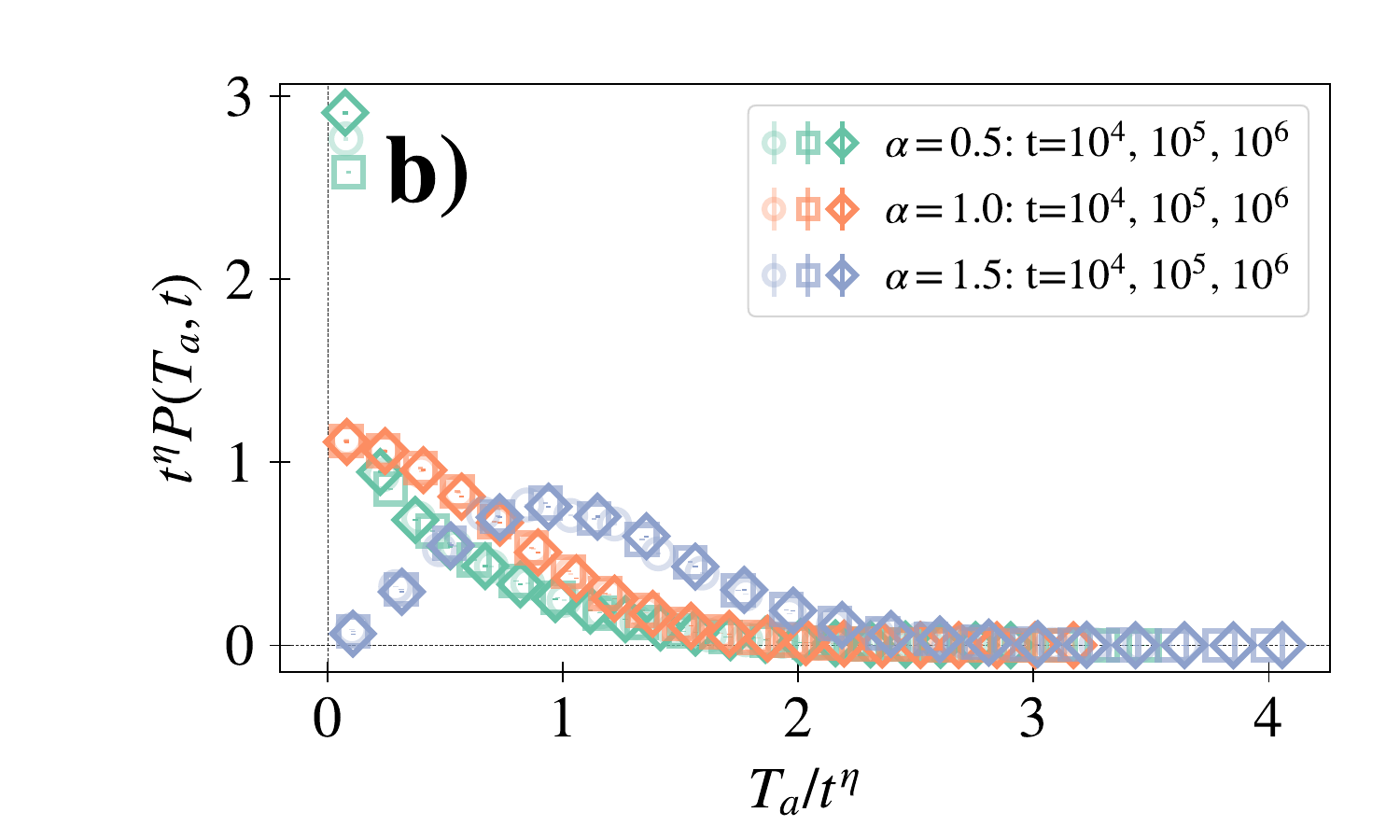}
    \caption{Scaling form for $T^+$ -a)- and $T_a$ -b)- for the SBM for three values of $\alpha=0.5,1,$ and $1.5$ in green, orange and blue symbols respectively. For each parameter value, we show three trajectory final times: $t=10^4,10^5,$ and $10^6$ in circles, squares, and diamonds, respectively. In panel b) $\eta=1-\alpha/2$. The other simulation parameters are $a=0.5$, $x_0=0$, $K=1$, time discretization of $dt=0.1$, and $N=10^5$ trajectories.}
    \label{fig:Scaling_SBM}
\end{figure}

From the results for the first and second moments of $T^+$ and $T_a$, we can derive the scaling form of the PDFs of these observables for the SBM and fBM in the long time limit since $\left\langle T^{+}(t)^{2}\right\rangle \sim\left\langle T^{+}(t)\right\rangle ^{2}$ and $\left\langle T_{a}(t)^{2}\right\rangle \sim\left\langle T_{a}(t)\right\rangle ^{2}$. 
Assume that in the long time limit the PDF of $T^+$ has the scaling form  $P(T^{+},t)\sim g_{+}(T^{+}/t^{\gamma})/t^{\eta}$. Then, 
$$
\left\langle T^{+}(t)\right\rangle \sim\int_{0}^{\infty}T^{+}P(T^{+},t)dT^{+}=t^{2\gamma-\eta}\int_{0}^{\infty}ug_{+}(u)du
$$
and
$$
\left\langle T^{+}(t)^2\right\rangle \sim\int_{0}^{\infty}T^{+2}P(T^{+},t)dT^{+}=t^{3\gamma-\eta}\int_{0}^{\infty}u^2g_{+}(u)du.
$$
Since $\left\langle T^{+}(t)\right\rangle \sim t$ and $\left\langle T^{+}(t)^{2}\right\rangle \sim t^{2}$ we have $\gamma=\eta=1$ and the scaling is 
\begin{eqnarray}
  P(T^{+},t)\sim\frac{1}{t}g_{+}\left(\frac{T^{+}}{t}\right),
  \label{sc1}
\end{eqnarray}
for both SBM and fBM. Note that arcsine law and the Lamperti distribution obey this scaling. Proceeding analogously with $T_a$ we get 
\begin{eqnarray}
    P(T_{a},t) \sim\frac{1}{t^{\eta}}g_\eta\left(\frac{T_{a}}{t^{\eta}}\right)
    \label{sc2}
\end{eqnarray}
where $\eta=1-\alpha/2$
for SBM and $\eta=1-H$ for fBM. Note also that in the Brownian limit ($\alpha=1$ and $H=1/2$) this scaling is in agreement with the half Gaussian distribution. 

\begin{figure}[!htbp]
    \centering
    \includegraphics[width=0.5\linewidth]{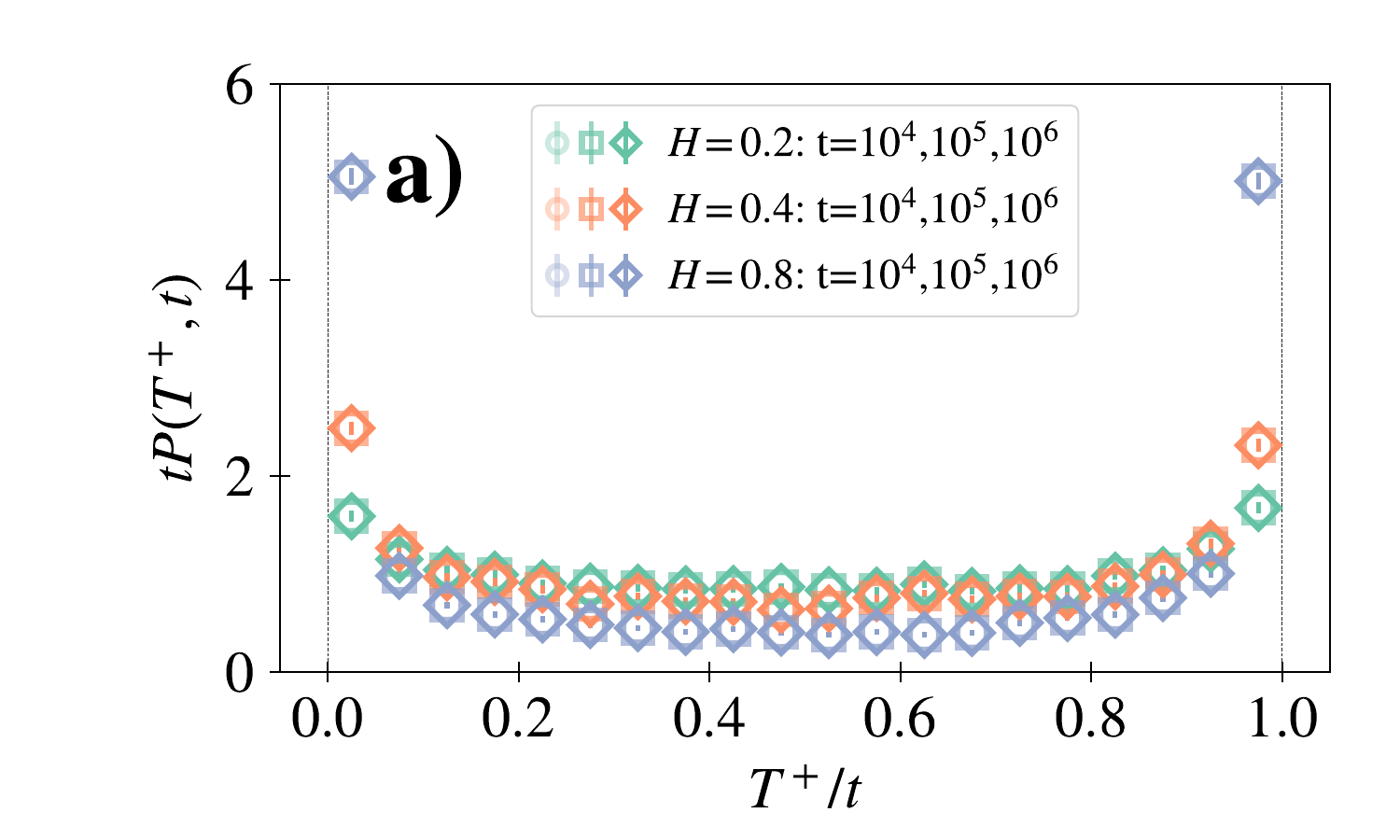}%
    \includegraphics[width=0.5\linewidth]{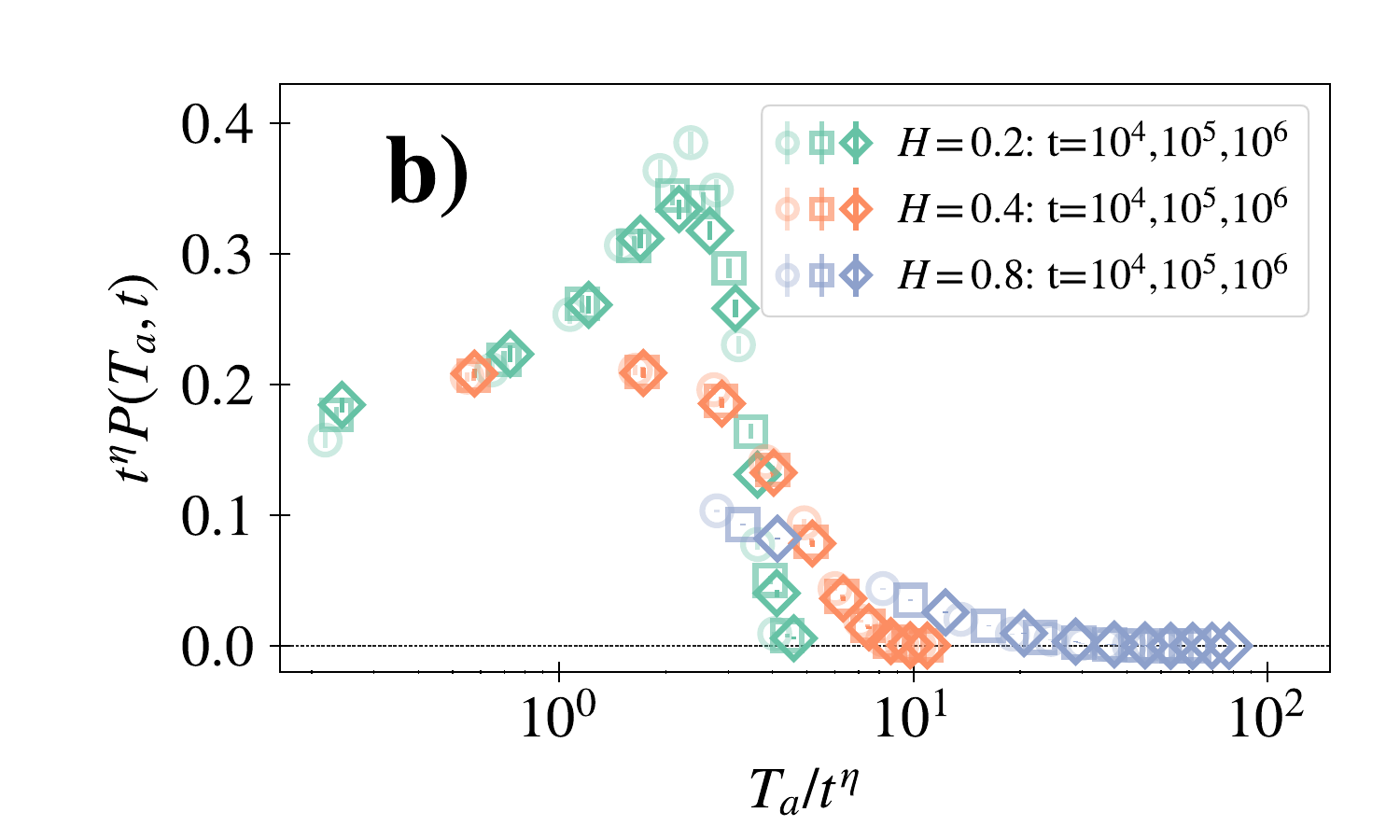}
    \caption{Scaling form for $T^+$ -a)- and $T_a$ -b)- for the fBM for three values of $H=0.2,0.4$ and $0.8$ in green, orange, and blue symbols respectively. For each parameter value, we show three trajectory final times: $t=10^4,10^5,$ and $10^6$ in circles, squares, and diamonds, respectively. In panel b) $\eta=1-H$. The other simulation parameters are  $a=2$, $x_0=0$, $D_H=1/2$, number of time points $M=2^{23}$, and $N=10^4$ trajectories.}
    \label{fig:Scaling_fBM}
\end{figure}

In Figs. \ref{fig:Scaling_SBM} and \ref{fig:Scaling_fBM} we check the scalings \eqref{sc1} (panels a)) and \eqref{sc2} (panels b)) with simulations for SBM and fBM respectively. The scaling functions $g_+(\cdot)$ and $g_\eta(\cdot)$ depend on the values of parameters $\alpha$ and $H$. We show that for fixed values of the parameters and different values of $t$ the data collapse on the same curve, proving the scalings.

Finally, from infinite ergodic theory we can obtain an interesting property for any observable that fulfills Eq. \eqref{eq:integrability} For the SBM, introducing  Eq. \eqref{cmsd} into  Eq. \eqref{quo2} we readily find
\begin{eqnarray}
    \xi = \lim_{t\to\infty}\frac{\left\langle \overline{\mathcal{O}}\right\rangle }{\left\langle \mathcal{O}\right\rangle }=\frac{2}{2-\alpha}.
    \label{quo3}
\end{eqnarray}
Analogously, for the fBM, using Eq. \eqref{C} and Eq. \eqref{quo2} we obtain
\begin{eqnarray}
    \xi = \lim_{t\to\infty}\frac{\left\langle \overline{\mathcal{O}}\right\rangle }{\left\langle \mathcal{O}\right\rangle }=\frac{1}{1-H}.
    \label{quo4}
\end{eqnarray}

\begin{figure}[!htbp]
    \centering
    \includegraphics[width=0.5\linewidth]{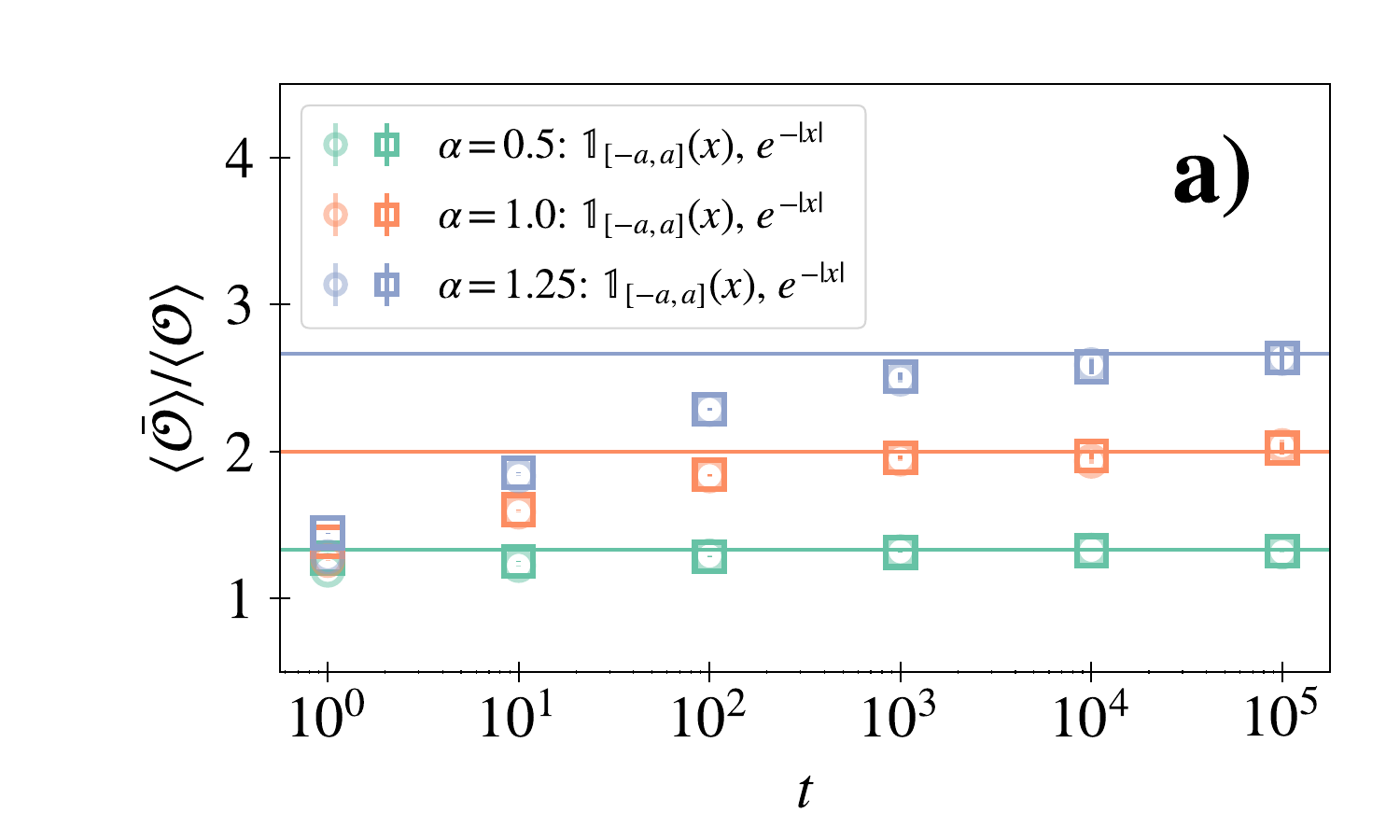}%
    \includegraphics[width=0.5\linewidth]{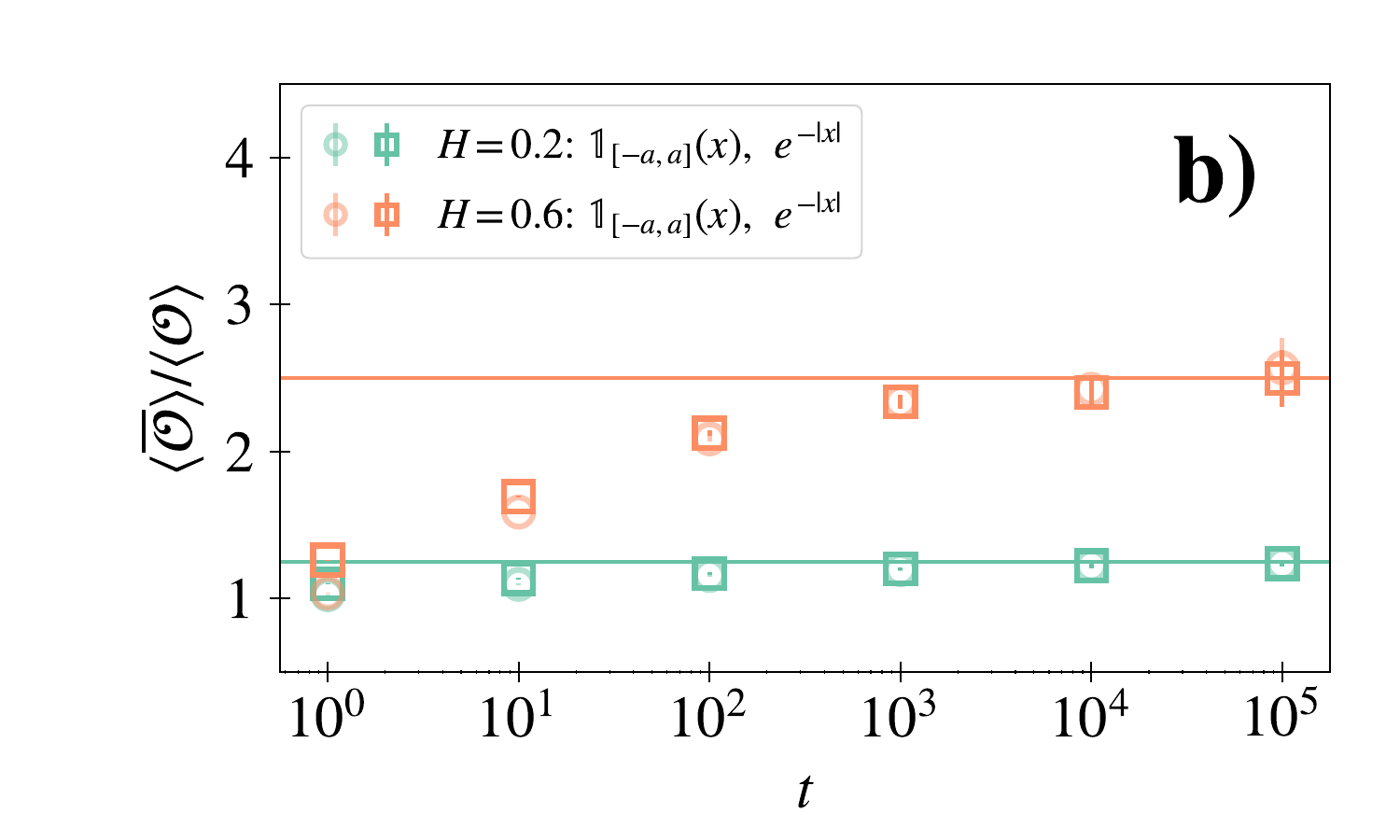}
    \caption{Quotient $\xi = \lim_{t\to\infty}\left\langle \overline{\mathcal{O}}\right\rangle /\left\langle \mathcal{O}\right\rangle$ for the observables  $\mathcal{O}(x)=\mathds{1}_{[-a,a]}(x)$ (circles) and $\mathcal{O}(x)=e^{-|x|}$ (squares) as function of time for both the SBM in panel a), and the fBM in panel b). The solid lines represent Eq. \eqref{quo3} and Eq. \eqref{quo4} respectively. The symbols are from numerical simulations. In panel a) the parameters used for the simulation are: $\alpha=0.5,1.0,$ and $1.25$, $N=7\cdot10^5$ trajectories, $x_0=0$, $K=1$, and time discretization of $dt=0.1$. In panel b) $H=0.2,$ and $0.6$$x_0=0$, $D_H=1/2$, number of time points $M=2^{22}$, and $N=2\cdot10^4$ trajectories. For both plots $a=2$.}
    \label{fig:quo}
\end{figure}

In Fig. \ref{fig:quo} we check this result with simulations for two specific observables:  $\mathcal{O}(x)=\mathds{1}_{[-a,a]}(x)$ and $\mathcal{O}(x)=e^{-|x|}$. In figure \ref{fig:quo} we have checked that $\xi$ converges to Eq. \eqref{quo3} and Eq. \eqref{quo4} for different observables. We can also see that the convergence is slower for larger values of $\alpha$ and $H$, that is, in the superdiffusive regime. In this regime both the number of trajectories and the total evolution time have to be large in order to see the scaling.

It is interesting to observe that SBM and fBM are microscopic models which give rise to anomalous diffusion in the macroscopic limit. Many studies have analyzed the microscopic differences between them \cite{Me14,Wa22}. Our results for the two first moments and the EB parameter of the occupation times also illustrate statistical differences between SBM and fBM.

\section{Conclusions}
In this paper, we have made use of the Kac formalism to derive the first two moments of positive stochastic functionals in terms of the one-time and two-time PDFs of the underlying random walk. This formalism also allows us to show that observables of stationary random walks are ergodic, i.e., their PDFs converge to a Dirac delta function in the long time limit. We have derived exact analytical expressions for the two first moments of functionals as the half occupation time and the occupation time in an interval, when the random walk is Gaussian. These expressions have been further simplified in the long time limit when the random walk is in addition isotropic.  We have shown that the expression for the first moments can be also derived from the infinite ergodic theory. We have explicitly considered some relevant examples of Gaussian random walks. For example, we have considered the case of a Brownian motion with time-dependent diffusivity, the SBM and the fBM. We compared the EB parameter for SBM and fBM with that from the ML distribution for occupation time in an interval. We have seen that the ML distribution is not in general the PDF of the occupation time in an interval for the SBM and fBM, but only for specific values of the exponents $\alpha$ and $H$. Nevertheless, our analytic expressions for the EB parameter agree well with the numerical simulations. Since the occupation times are such that the second moment scales as the square of the first moment we can infer a simple scaling form of their corresponding PDFs in the long time limit. Our theoretical results have been tested with Montecarlo simulation exhibiting an excellent agreement. 

The methodology used in this paper could be useful to study the statistical properties of functionals of other stochastic processes when the solution of the FK equation is unknown or very difficult to solve analytically. 
\textcolor{black}{In this work, we have considered that the underlying process is Gaussian, which allows us to calculate the one-time and two-time PDFs directly from the characteristic functional. This method could also be applied to non-Gaussian processes, such as L\'evy flights, Poisson random walk, dichotomous random walk ..., for which the characteristic functionals are also known. However, even if the characteristic functional is not available, the two first moments and the EB parameter can be computed if the one-time and two-time PDFs are known.}

\section*{Acknowledgements}
The authors acknowledge the financial support of the Ministerio de Ciencia e Innovaci\'on (Spanish government) under
Grant No. PID2021-122893NB-C22.

\appendix

\section{Computation of the two first moments} \label{app:moments}
Let us check that the first two moments can be obtained from \eqref{zn} and \eqref{Gn}. On setting $n=0$ into \eqref{Gn}
\begin{eqnarray}
G_{1}(x,t)=\int_{0}^{t}d\tau_{1}\int_{-\infty}^{\infty}dx_{1}U(x_{1})P(x_{1},\tau_{1})P(x,t|x_{1},\tau_{1})
    \label{G1}
\end{eqnarray}
and from \eqref{zn}
\begin{eqnarray}
\left\langle Z(t)\right\rangle &=&\int_{-\infty}^{\infty}G_{1}(x,t)dx=\int_{0}^{t}d\tau_{1}\int_{-\infty}^{\infty}dx_{1}U(x_{1})P(x_{1},\tau_{1})\int_{-\infty}^{\infty}dxP(x,t|x_{1},\tau_{1})\nonumber\\
&=&\int_{0}^{t}d\tau_{1}\int_{-\infty}^{\infty}dx_{1}U(x_{1})P(x_{1},\tau_{1})
\label{z1}
\end{eqnarray}
where $\int_{-\infty}^{\infty}dxP(x,t|x_{1},\tau_{1})=1$ by normalization. Analogously, on setting $n=1$ into \eqref{Gn}
\begin{eqnarray}
&G_{2}&(x,t)= \nonumber\\
&=& \int_{0}^{t}d\tau_{2}\int_{-\infty}^{\infty}dx_{2}U(x_{2})G_{1}(x_{2},\tau_{2})P(x,t|x_{2},\tau_{2}) \nonumber\\
&=&\int_{0}^{t}d\tau_{2}\int_{-\infty}^{\infty}dx_{2}U(x_{2})P(x,t|x_{2},\tau_{2})\int_{0}^{\tau_{2}}d\tau_{1}\int_{-\infty}^{\infty}dx_{1}U(x_{1})P(x_{1},\tau_{1})P(x_{2},\tau_{2}|x_{1},\tau_{1})
    \label{G2}
\end{eqnarray}
where we have considered $G_1(x,t)$ from \eqref{G1}. Finally, from \eqref{zn} and \eqref{G2}
\begin{eqnarray}
\left\langle Z(t)^{2}\right\rangle &=&2!\int_{-\infty}^{\infty}G_{2}(x,t)dx\nonumber\\
&=&2!\int_{0}^{t}d\tau_{2}\int_{0}^{\tau_{2}}d\tau_{1}\int_{-\infty}^{\infty}dx_{2}U(x_{2})\int_{-\infty}^{\infty}dx_{1}U(x_{1})P(x_{2},\tau_{2};x_{1},\tau_{1}).
    \label{z22}
\end{eqnarray}
Indeed, Eqs. \eqref{z1} and \eqref{z22} are equal to \eqref{Z} and \eqref{z21} as expected.

\section{Derivation of Eq. \eqref{tfU1}} \label{app:eq48}
Consider the following integral:
\begin{align}
    I(k)=\int_{-\infty}^{\infty}dx\,e^{-ikx}\theta(x)=\int_0^\infty dx\cos(kx)-i\int_0^\infty dx\sin(kx). \label{Iki1}
\end{align}
The last two integrals are not convergent in the Riemann or Lebesgue sense, so $I(k)$ either. However, it is possible to express $I(k)$ in terms of distributions. To do so, consider
\begin{align}
    J(k)=\int_{-\infty}^{\infty}dx\,e^{-ikx}\theta(x)e^{-\varepsilon x}=\int_{\textcolor{black}{0}}^\infty dx\,e^{-i(k+\varepsilon)x}, \label{Jki1}
\end{align}
being $\varepsilon$ a positive constant. Now, $J(k)$ is integrable, since $e^{-\varepsilon x}\to0$ as $x\to\infty$. Therefore, \eqref{Jki1} becomes
\begin{align}
    J(k)=\frac{1}{ik+\varepsilon}.
\end{align}
On one hand, note that $\lim_{\varepsilon\to0^+}J(k)=I(k)$. On the other hand, the Sokhotski-Plemelj identity states that
\begin{align}
    \lim_{\varepsilon\to0^+}\frac{1}{k\pm i\varepsilon}=\mathcal{P}\left(\frac{1}{k}\right)\mp i\pi\delta(k),
\end{align}
where $\mathcal{P}$ denotes the Cauchy's principal value. In consequence, \eqref{Iki1} reads
\begin{align}
    I(k)=-i\left[\mathcal{P}\left(\frac{1}{k}\right)+i\pi\delta(k)\right]=-\frac{i}{k}+\pi\delta(k).
\end{align}

\section{Derivation of Eq. \eqref{t2g}} \label{app:eqt2g}
Let us first compute the integrals in Eq. \eqref{z2f} over $k_1$ and $k_2$ separately using \eqref{tfU1}. Hence, 
\begin{eqnarray}
    \int_{-\infty}^{\infty}&dk_{1}&\int_{-\infty}^{\infty}dk_{2}\tilde{P}(k_{2},\tau_{2};k_{1},\tau_{1})\tilde{U}(-k_{1})\tilde{U}(-k_{2})=\pi^{2}-\int_{-\infty}^{\infty}\frac{dk_{1}}{k_{1}}\int_{-\infty}^{\infty}\frac{dk_{2}}{k_{2}}\tilde{P}(k_{2},\tau_{2};k_{1},\tau_{1})\nonumber\\
    &-&i\pi\int_{-\infty}^{\infty}\frac{dk_{1}}{k_{1}}\tilde{P}(k_{1},\tau_{1})
    -i\pi\int_{-\infty}^{\infty}\frac{dk_{2}}{k_{2}}\tilde{P}(k_{2},\tau_{2}).
    \label{intk}
\end{eqnarray}
The second and third integrals of the right hand side in the above expression are actually the same integral and using \eqref{cfot} one finds
\begin{eqnarray}
\int_{-\infty}^{\infty}\frac{dk_{i}}{k_{i}}\tilde{P}(k_{i},\tau_{i})=i\pi\;\textrm{erf}\left(\frac{\left\langle x(\tau_{i})\right\rangle }{\sqrt{2C(\tau_{i},\tau_{i})}}\right)
    \label{intk1}
\end{eqnarray}
with $i=1,2$. On the other hand, the first integral can be computed using \eqref{cftt}. To this end we define 
\begin{eqnarray}
    I_{1}(C(\tau_{1},\tau_{2}))=\int_{-\infty}^{\infty}\frac{dk_{1}}{k_{1}}\int_{-\infty}^{\infty}\frac{dk_{2}}{k_{2}}\tilde{P}(k_{2},\tau_{2};k_{1},\tau_{1})
    \label{I1i}
\end{eqnarray}
and note that
$$
-\frac{\partial I_{1}}{\partial C(\tau_{1},\tau_{2})}=\int_{-\infty}^{\infty}dk_{1}\int_{-\infty}^{\infty}dk_{2}\tilde{P}(k_{2},\tau_{2};k_{1},\tau_{1}).
$$
Computing first the integral over $k_2$ we find
\begin{eqnarray*}
    \int_{-\infty}^{\infty}\tilde{P}(k_{2},\tau_{2};k_{1},\tau_{1})dk_{2}&=&e^{ik\left\langle x(\tau_{1})\right\rangle -\frac{k_{1}^{2}}{2}C(\tau_{1},\tau_{1})}\int_{-\infty}^{\infty}e^{ik\left\langle x(\tau_{2})\right\rangle -\frac{k_{2}^{2}}{2}C(\tau_{2},\tau_{2})-k_{1}k_{2}C(\tau_{1},\tau_{2})}dk_{2}\\&=&\sqrt{\frac{2\pi}{C(\tau_{2},\tau_{2})}}e^{ik\left\langle x(\tau_{1})\right\rangle -\frac{k_{1}^{2}}{2}C(\tau_{1},\tau_{1})}e^{\frac{(k_{1}C(\tau_{1},\tau_{2})-i\left\langle x(\tau_{2})\right\rangle )^{2}}{2C(\tau_{2},\tau_{2})}},
\end{eqnarray*}
where we have made use of the result
\begin{eqnarray}
    \int_{-\infty}^{\infty}e^{-\beta z^{2}-\gamma z}\cos(bz)dz=\frac{1}{2}\sqrt{\frac{\pi}{\beta}}\left[e^{\frac{(\gamma-ib)^{2}}{4\beta}}+e^{\frac{(\gamma+ib)^{2}}{4\beta}}\right]
    \label{aux}
\end{eqnarray}
with $b=0$. Rearranging the factors, the integral over $k_1$ can be written as
\begin{eqnarray*}
  \int_{-\infty}^{\infty}dk_{1}\int_{-\infty}^{\infty}dk_{2}\tilde{P}(k_{2},\tau_{2};k_{1},\tau_{1})&=&\sqrt{\frac{2\pi}{C(\tau_{2},\tau_{2})}}e^{-\frac{\left\langle x(\tau_{2})\right\rangle ^{2}}{2C(\tau_{2},\tau_{2})}}\\
  &\times&\int_{-\infty}^{\infty}dk_{1}\cos\left[k_{1}\left(\left\langle x(\tau_{1})\right\rangle -\left\langle x(\tau_{2})\right\rangle \frac{C(\tau_{1},\tau_{2})}{C(\tau_{2},\tau_{2})}\right)\right]e^{-k_{1}^{2}\frac{\Delta\left(\tau_{1},\tau_{2}\right)}{2C(\tau_{2},\tau_{2})}},
\end{eqnarray*}
where $\Delta\left(\tau_{1},\tau_{2}\right)=C(\tau_{1},\tau_{1})C(\tau_{2},\tau_{2})-C(\tau_{1},\tau_{2})^{2}$. The integral of the right hand side can be solved with the help of \eqref{aux} setting $\gamma=0$, so that
$$
\int_{-\infty}^{\infty}dk_{1}\int_{-\infty}^{\infty}dk_{2}\tilde{P}(k_{2},\tau_{2};k_{1},\tau_{1})=\frac{2\pi}{\sqrt{\Delta\left(\tau_{1},\tau_{2}\right)}}e^{-\frac{\left\langle x(\tau_{2})\right\rangle ^{2}}{2C(\tau_{2},\tau_{2})}}e^{-\frac{\left[\left\langle x(\tau_{1})\right\rangle -\left\langle x(\tau_{2})\right\rangle \frac{C(\tau_{1},\tau_{2})}{C(\tau_{2},\tau_{2})}\right]^{2}}{2\Delta\left(\tau_{1},\tau_{2}\right)}C(\tau_{2},\tau_{2})}.
$$
To find $I_{1}(C(\tau_{1},\tau_{2}))$ we have to integrate the previous expression over $C(\tau_{1},\tau_{2})$. The integral cannot be solved but can be expressed as
\begin{eqnarray}
    I_{1}(C(\tau_{1},\tau_{2}))=-F(y,\xi_{1},\xi_{2})+\phi
    \label{I1gen}
\end{eqnarray}
where $\phi$ is an integration constant that may depend on the other parameters and
$$
F(y,\xi_{1},\xi_{2})=2\pi\int\frac{dy}{\sqrt{1-y^{2}}}\exp\left[-\frac{\xi_{1}^{2}+\xi_{2}^{2}-2y\xi_{1}\xi_{2}}{2(1-y^{2})}\right]
$$
with
$$
y=\frac{C(\tau_{1},\tau_{2})}{\sqrt{C(\tau_{1},\tau_{1})C(\tau_{2},\tau_{2})}},\quad \xi_{i}=\frac{\left\langle x(\tau_{i})\right\rangle }{\sqrt{C(\tau_{i},\tau_{i})}}
$$
$i=1,2$. To find $\phi$ we consider $C(\tau_{1},\tau_{2})=0$ in Eq. \eqref{I1i} and observe that in this case $\tilde{P}(k_{2},\tau_{2};k_{1},\tau_{1})=\tilde{P}(k_{2},\tau_{2})\tilde{P}(k_{1},\tau_{1})$. This means that
$$
I_{1}(0)=\int_{-\infty}^{\infty}\frac{dk_{1}}{k_{1}}\tilde{P}(k_{1},\tau_{1})\int_{-\infty}^{\infty}\frac{dk_{2}}{k_{2}}\tilde{P}(k_{2},\tau_{2})=-\pi^2\;\textrm{erf}\left(\frac{\left\langle x(\tau_{1})\right\rangle }{\sqrt{2C(\tau_{1},\tau_{1})}}\right)\textrm{erf}\left(\frac{\left\langle x(\tau_{2})\right\rangle }{\sqrt{2C(\tau_{2},\tau_{2})}}\right)
$$
where we made use of \eqref{intk1}. On the other hand, setting $C(\tau_{1},\tau_{2})=0$ in Eq. \eqref{I1gen} we have $I_1(0)=-F(0,\xi_{1},\xi_{2})+\phi$, which fixes the expression for $\phi$. Finally, from \eqref{I1gen} and \eqref{I1i} we obtain
\begin{eqnarray*}
    \int_{-\infty}^{\infty}\frac{dk_{1}}{k_{1}}\int_{-\infty}^{\infty}\frac{dk_{2}}{k_{2}}\tilde{P}(k_{2},\tau_{2};k_{1},\tau_{1})&=&-\pi^2\;\textrm{erf}\left(\frac{\left\langle x(\tau_{1})\right\rangle }{\sqrt{2C(\tau_{1},\tau_{1})}}\right)\textrm{erf}\left(\frac{\left\langle x(\tau_{2})\right\rangle }{\sqrt{2C(\tau_{2},\tau_{2})}}\right)\\
    &-&2\pi\int_{0}^{\frac{C(\tau_{1},\tau_{2})}{\sqrt{C(\tau_{1},\tau_{1})C(\tau_{2},\tau_{2})}}}\frac{dz}{\sqrt{1-z^{2}}}\exp\left[-\frac{\xi_{1}^{2}+\xi_{2}^{2}-2z\xi_{1}\xi_{2}}{2(1-z^{2})}\right].
\end{eqnarray*}
Finally, introducing the above expression 
and \eqref{intk1} into \eqref{intk}  we obtain \eqref{t2g}
 using \eqref{z2f}.

\section{Derivation of Eqs. \eqref{Ta1} and \eqref{Ta2ex}} \label{app:Ta}

We detail the derivation of the general expression for $\langle T_a(t)\rangle$ and $\langle T_a^2(t) \rangle$ in terms of $\langle x(t) \rangle$ and autocorrelation of the Gaussian process: Eqs. \eqref{Ta1} and \eqref{Ta2ex}. Let us start with the first moment. To compute the integral over $k$ in Eq. \eqref{ta1g} we need to introduce a parametric derivative as follows. Define the integral
\begin{eqnarray}
I_2(a)=\int_{-\infty}^{\infty}\frac{\sin(ka)}{k}e^{ik\left\langle x(\tau)\right\rangle -\frac{k^{2}}{2}C(\tau,\tau)}dk.
\label{I1}
\end{eqnarray}
Its derivative with respect to $a$ can be found using \eqref{aux} so that
$$
\frac{\partial I_2(a)}{\partial a}=\int_{-\infty}^{\infty}\cos(ka)e^{ik\left\langle x(\tau)\right\rangle -\frac{k^{2}}{2}C(\tau,\tau)}dk=\frac{1}{2}\sqrt{\frac{2\pi}{C(\tau,\tau)}}\left[e^{-\frac{(a+\left\langle x(\tau)\right\rangle )^{2}}{2C(\tau,\tau)}}+e^{-\frac{(a-\left\langle x(\tau)\right\rangle )^{2}}{2C(\tau,\tau)}}\right].
$$
Now, integrating the previous equation over $a$ we find
\begin{eqnarray}
  I_2(a)=\frac{\pi}{2}\left[\textrm{erf}\left(\frac{a-\left\langle x(\tau)\right\rangle }{\sqrt{2C(\tau,\tau)}}\right)+\textrm{erf}\left(\frac{a+\left\langle x(\tau)\right\rangle }{\sqrt{2C(\tau,\tau)}}\right)\right]
  \label{I12}
\end{eqnarray}
where the integration constant is zero provided that $I_2(a=0)=0$. Finally, Eq. \eqref{ta1g} has the form
\begin{eqnarray}
    \left\langle T_{a}(t)\right\rangle =\frac{1}{2}\int_{0}^{t}\left[\textrm{erf}\left(\frac{a-\left\langle x(\tau)\right\rangle }{\sqrt{2C(\tau,\tau)}}\right)+\textrm{erf}\left(\frac{a+\left\langle x(\tau)\right\rangle }{\sqrt{2C(\tau,\tau)}}\right)\right]d\tau \nonumber
\end{eqnarray}
which is \eqref{Ta1}. 

The second moment $\left\langle T_{a}(t)^2\right\rangle$ is found from \eqref{ta2i}. We insert \eqref{cftt} into \eqref{ta2i} and perform the integrals but only one integral with respect to $k_2$ or $k_1$ can be solved explicitly. To compute the integral over $k_2$ we define the integral
$$
I_{2}(a,k_1,\tau_1,\tau_2)=\int_{-\infty}^{\infty}dk_{2}\frac{\sin(k_{2}a)}{k_{2}}e^{ik_{2}\left\langle x(\tau_{2})\right\rangle -k_{1}k_{2}C(\tau_{1},\tau_{2})-\frac{k_{2}^{2}}{2}C(\tau_{2},\tau_{2})}.
$$
Its derivative with respect to $a$ can be found explicitly
\begin{eqnarray*}
    \frac{\partial I_{2}(a,k_1,\tau_1,\tau_2)}{\partial a}&=&\int_{-\infty}^{\infty}dk_{2}\cos(k_{2}a)e^{ik_{2}\left\langle x(\tau_{2})\right\rangle -k_{1}k_{2}C(\tau_{1},\tau_{2})-\frac{k_{2}^{2}}{2}C(\tau_{2},\tau_{2})}\nonumber\\
    &=&\sqrt{\frac{\pi}{2C(\tau_{2},\tau_{2})}}\left[e^{-\frac{\left(a+\left\langle x(\tau_{2})\right\rangle +ik_{1}C(\tau_{1},\tau_{2})\right)^{2}}{2C(\tau_{2},\tau_{2})}}+e^{-\frac{\left(a-\left\langle x(\tau_{2})\right\rangle -ik_{1}C(\tau_{1},\tau_{2})\right)^{2}}{2C(\tau_{2},\tau_{2})}}\right].
\end{eqnarray*}
Integrating over $a$ we find 
$$
I_{2}(a,k_1,\tau_1,\tau_2)=\frac{\pi}{2}\left[\textrm{erf}\left(\frac{a-\left\langle x(\tau_{2})\right\rangle -ik_{1}C(\tau_{1},\tau_{2})}{\sqrt{2C(\tau_{2},\tau_{2})}}\right)+\textrm{erf}\left(\frac{a+\left\langle x(\tau_{2})\right\rangle +ik_{1}C(\tau_{1},\tau_{2})}{\sqrt{2C(\tau_{2},\tau_{2})}}\right)\right]
$$
which combined with \eqref{ta2i} yields
\begin{eqnarray*}
    \left\langle T_{a}(t)^{2}\right\rangle =\frac{2}{\pi^{2}}\int_{0}^{t}d\tau_{2}\int_{0}^{\tau_{2}}d\tau_{1}\int_{-\infty}^{\infty}dk_{1}\frac{\sin(k_{1}a)}{k_{1}}e^{ik_{1}\left\langle x(\tau_{1})\right\rangle -\frac{k_{1}^{2}}{2}C(\tau_{1},\tau_{1})}I_{2}(a,k_{1},\tau_{1},\tau_{2}).
\end{eqnarray*}
which is Eq. \eqref{Ta2ex}.

\section{Derivation of Eq. \eqref{eq:SBM_TPl2}}
Since $C(\tau_1,\tau_1)C(\tau_2,\tau_2)>C(\tau_1,\tau_2)^2$, we can use the trigonometric identity
\begin{align}
    \arctan\left(\sqrt{\frac{C(\tau_{2},\tau_{2})C(\tau_{1},\tau_{1})}{C(\tau_{1},\tau_{2})^{2}}-1}\right)=\frac{\pi}{2}-\arcsin\left(\frac{C(\tau_1,\tau_2)}{\sqrt{C(\tau_{1},\tau_{1})C(\tau_2,\tau_2)}}\right).
\end{align}
Therefore, \eqref{t1t2} reduces to
\begin{align}
    \left\langle T^{+}(t)^{2}\right\rangle =\frac{t^{2}}{4}+\frac{1}{\pi}\int_{0}^{t}d\tau_{2}\int_{0}^{\tau_{2}}d\tau_{1}\arcsin\left(\frac{C(\tau_1,\tau_2)}{\sqrt{C(\tau_{1},\tau_{1})C(\tau_2,\tau_2)}}\right). \label{t+2nou}
\end{align}
Using \eqref{Dsbm}, \eqref{t+2nou} reads
\begin{align}
\left\langle T^{+}(t)^{2}\right\rangle &=\frac{t^{2}}{4}+\frac{1}{\pi}\int_{0}^{t}d\tau_{2}\int_{0}^{\tau_{2}}d\tau_{1}\arcsin\left(\sqrt{\frac{\tau_{1}^{\alpha}}{\tau_{1}^{\alpha}}}\right).
\label{tm20}
\end{align}
Considering the change of variables $u=\sqrt{\frac{\tau_1^\alpha}{\tau_2^\alpha}}$, \eqref{tm20} gives
\begin{align}
    \langle T^+(t)^2\rangle=\frac{t^2}{4}+\frac{t^2}{\pi\alpha}\int_0^1 du\,u^{\frac{2}{\alpha}-1}\arcsin(u).\label{tm2'}
\end{align}
Taking $u'=\arcsin(u)$ and $dv'=u^{\frac{2}{\alpha}-1}du$ and integrating by parts, \eqref{tm2'} becomes
\begin{align}
    \langle T^+(t)^2\rangle=\frac{t^2}{2}-\frac{t^2}{2\pi}\int_0^1du\, u^{\frac{2}{\alpha}}(1-u^2)^{-\frac{1}{2}}.\label{tm2''}
\end{align}
Consider now the change of variables $w=u^2$. Then, \eqref{tm2''} turns into
\begin{align}
    \langle T^+(t)^2\rangle=\frac{t^2}{2}-\frac{t^2}{4\pi}\int_0^1 dw\,w^{\frac{1}{\alpha}+\frac{1}{2}-1}(1-w)^{\frac{1}{2}-1}. \label{tm2'''}
\end{align}
The beta function is defined as $B(a,b)=\int_0^1 dw\,w^{a-1} (1-w)^{b-1}$,
where $a$ and $b$ are real entries. Then, the  integral in \eqref{tm2'''} is equal to $B\left(\frac{1}{\alpha}+\frac{1}{2},\frac{1}{2}\right)$. Also, the relation between the beta and the gamma function is given as $B(a,b)=\frac{\Gamma(a)\Gamma(b)}{\Gamma(a+b)}$. Thus, \eqref{tm2'''} finally looks as
\begin{align}
    \langle T^+(t)^2\rangle=\frac{t^2}{2}\left[1-\frac{1}{2\sqrt{\pi}}\frac{\Gamma\left(\frac{1}{\alpha}+\frac{1}{2}\right)}{\Gamma\left(1+\frac{1}{\alpha}\right)}\right]. \label{aquesta}
\end{align}

\section{Derivation of Eq. \eqref{eq:SBM_Ta2}}
After combining \eqref{Ta22} and \eqref{Dsbm}, we get 
\begin{align}
    \langle T_a(t)^2\rangle&\mathop{\simeq}\limits_{t \to \infty}\frac{2a^2t^2}{\pi K}\int_0^1 dv\,v\int_0^1 du\,(uvt)^{-\frac{\alpha}{2}}\left((vt)^\alpha-(uvt)^\alpha\right)^{-\frac{1}{2}}\nonumber\\
           &=\frac{2a^{2}t^{2-\alpha}}{\pi K(2-\alpha)}
\int_{0}^{1}du\,u^{-\frac{\alpha}{2}}
           \left(1-u^{\alpha}\right)^{-\frac{1}{2}}. \label{Ta2SBM'}
\end{align}
\noindent Now, considering the change of variables $w=u^\alpha$, \eqref{Ta2SBM'} turns into
\begin{align}
     \langle T_a(t)^2\rangle\mathop{\simeq}\limits_{t \to \infty}\frac{2a^2t^{2-\alpha}}{\alpha(2-\alpha)\pi K}\int_0^1dw\,w^{\frac{1}{\alpha}-\frac{3}{2}}(1-w)^{-\frac{1}{2}}. \label{Ta2SBM''}
\end{align}
The integral above is equal to $B\left(\frac{1}{\alpha}-\frac{1}{2},\frac{1}{2}\right)$. Provided that $B(a,b)=\frac{\Gamma(a)\Gamma(b)}{\Gamma(a+b)}$, \eqref{Ta2SBM''} reads
\begin{align}
    \langle T_{a}(t)^{2}\rangle \mathop{\simeq}\limits_{t \to \infty}\frac{2a^{2}}{K\sqrt{\pi}}\frac{\Gamma\left(\frac{1}{\alpha}-\frac{1}{2}\right)}{\alpha(2-\alpha)\Gamma\left(\frac{1}{\alpha}\right)}t^{2-\alpha}\,\,.\label{Ta2SBM'''}
\end{align}

\section{Derivation of Eq. \eqref{eq:fbmTp2}}

\noindent After introducing \eqref{C} in \eqref{t1t2}, we get that
\begin{align}
    \langle T^+(t)^2\rangle=\frac{t^2}{2}-\frac{1}{\pi}\int_0^t d\tau_2\int_0^{\tau_2}d\tau_1\arctan\left(\sqrt{\frac{4\tau_1^{2H}\tau_2^{2H}}{(\tau_1^{2H}+\tau_2^{2H}-(\tau_2-\tau_1)^{2H})^2}-1}\right). \label{t+2fbm1}
\end{align}
Since $0\leq\tau_2\leq t$, we can parametrize $\tau_2$ as $\tau_2=st$, with $s\in[0,1]$. Similarly, $\tau_1=u\tau_2$, with $u\in[0,1]$. Then, $d\tau_2\,d\tau_1=st^2\,ds\,du$.
Therefore, \eqref{t+2fbm1} becomes
\begin{align}
    \langle T^{+}(t)^{2}\rangle
   = \frac{t^{2}}{2}
     - \frac{t^{2}}{\pi}
       \int_{0}^{1} s\,ds
       \int_{0}^{1} du\,
       \arctan\!\left(
         \sqrt{ \frac{4\,u^{2H}}
                      {\bigl(u^{2H}+1-(1-u)^{2H}\bigr)^{2}}
                - 1 }
       \right),
\end{align}
which yields
\begin{align}
    \langle T^+(t)^2\rangle=\frac{t^2}{2}\left[1-\frac{1}{\pi}\int_{0}^{1} du\,
       \arctan\left(
         \frac{\sqrt{4u^{2H}-(1+u^{2H}-(1-u)^{2H})^{2}}}
              {1+u^{2H}-(1-u)^{2H}}
       \right)\right].
\end{align}

\bibliography{main}

@article{GoLu01,
author = {Godreche, C. and Luck, Jean-Marc},
year = {2001},
month = {11},
pages = {489–524},
title = {Statistics of the Occupation Time of Renewal Processes},
volume = {104},
journal = {Journal of Statistical Physics},
doi = {10.1023/A:1010364003250}
}

@article{sabhapandit2006statistical,
  title={Statistical properties of functionals of the paths of a particle diffusing in a one-dimensional random potential},
  author={Sabhapandit, Sanjib and Majumdar, Satya N and Comtet, Alain},
  journal={Physical Review E},
  volume={73},
  number={5},
  pages={051102},
  year={2006},
  publisher={APS},
  doi = {10.1103/PhysRevE.73.051102},
  url = {https://link.aps.org/doi/10.1103/PhysRevE.73.051102}
}

@article{Ma05,
  TITLE = {{Brownian Functionals in Physics and Computer Science}},
  AUTHOR = {Majumdar, Satya N.},
  URL = {https://hal.archives-ouvertes.fr/hal-00165789},
  JOURNAL = {{Current Science}},
  PUBLISHER = {{Indian Academy of Sciences}},
  VOLUME = {89},
  PAGES = {2076},
  YEAR = {2005},
  HAL_ID = {hal-00165789},
  HAL_VERSION = {v1},
}

@article{Kac49,
	author = {M. Kac},
	doi = {10.1090/s0002-9947-1949-0027960-x},
	journal = {Transactions of the American Mathematical Society},
	number = {1},
	pages = {1--13},
	publisher = {American Mathematical Society ({AMS})},
	title = {On distributions of certain Wiener functionals},
	url = {https://doi.org/10.1090%2Fs0002-9947-1949-0027960-x},
	volume = {65},
	year = 1949}

@article{Ca10,
	author = {Carmi, Shai and Turgeman, Lior and Barkai, Eli},
	date = {2010/12/01},
	doi = {10.1007/s10955-010-0086-6},
	id = {Carmi2010},
	isbn = {1572-9613},
	journal = {Journal of Statistical Physics},
	number = {6},
	pages = {1071--1092},
	title = {On Distributions of Functionals of Anomalous Diffusion Paths},
	url = {https://doi.org/10.1007/s10955-010-0086-6},
	volume = {141},
	year = {2010}}

@article{Le40,
author = {L\'evy, Paul},
journal = {Compositio Mathematica},
pages = {283-339},
publisher = {Johnson Reprint Corporation},
title = {Sur certains processus stochastiques homog\`enes},
url = {http://eudml.org/doc/88744},
volume = {7},
year = {1940},
}

@article{La58,
	author = {John Lamperti},
	doi = {10.1090/s0002-9947-1958-0094863-x},
	journal = {Transactions of the American Mathematical Society},
	number = {2},
	pages = {380--387},
	publisher = {American Mathematical Society ({AMS})},
	title = {An occupation time theorem for a class of stochastic processes},
	url = {https://doi.org/10.1090%2Fs0002-9947-1958-0094863-x},
	volume = {88},
	year = 1958}

@article{Ba06,
	author = {Barkai, E. },
	date = {2006/05/01},
	doi = {10.1007/s10955-006-9109-8},
	id = {Barkai2006},
	isbn = {1572-9613},
	journal = {Journal of Statistical Physics},
	number = {4},
	pages = {883--907},
	title = {Residence Time Statistics for Normal and Fractional Diffusion in a Force Field},
	url = {https://doi.org/10.1007/s10955-006-9109-8},
	volume = {123},
	year = {2006}}

@article{TuCaBa09,
  title = {Fractional Feynman-Kac Equation for Non-Brownian Functionals},
  author = {Turgeman, Lior and Carmi, Shai and Barkai, Eli},
  journal = {Physical Review Letters},
  volume = {103},
  issue = {19},
  pages = {190201},
  numpages = {4},
  year = {2009},
  month = {Nov},
  publisher = {American Physical Society},
  doi = {10.1103/PhysRevLett.103.190201},
  url = {https://link.aps.org/doi/10.1103/PhysRevLett.103.190201}
}

@article{Me14,
	author = {Metzler, Ralf and Jeon, Jae-Hyung and Cherstvy, Andrey G. and Barkai, Eli},
	doi = {10.1039/C4CP03465A},
	issue = {44},
	journal = {Physical Chemistry Chemical Physics},
	pages = {24128-24164},
	publisher = {The Royal Society of Chemistry},
	title = {Anomalous diffusion models and their properties: non-stationarity{,} non-ergodicity{,} and ageing at the centenary of single particle tracking},
	url = {http://dx.doi.org/10.1039/C4CP03465A},
	volume = {16},
	year = {2014}}

@article{ChMe15,
	author = {Andrey G Cherstvy and Ralf Metzler},
	doi = {10.1088/1742-5468/2015/05/P05010},
	journal = {Journal of Statistical Mechanics: Theory and Experiment},
	month = {may},
	number = {5},
	pages = {P05010},
	publisher = {IOP Publishing and SISSA},
	title = {Ergodicity breaking, ageing, and confinement in generalized diffusion processes with position and time dependent diffusivity},
	url = {https://dx.doi.org/10.1088/1742-5468/2015/05/P05010},
	volume = {2015},
	year = {2015}}

@article{Sa15,
	author = {Hadiseh Safdari and Andrey G Cherstvy and Aleksei V Chechkin and Felix Thiel and Igor M Sokolov and Ralf Metzler},
	doi = {10.1088/1751-8113/48/37/375002},
	journal = {Journal of Physics A: Mathematical and Theoretical},
	month = {aug},
	number = {37},
	pages = {375002},
	publisher = {IOP Publishing},
	title = {Quantifying the non-ergodicity of scaled Brownian motion},
	url = {https://dx.doi.org/10.1088/1751-8113/48/37/375002},
	volume = {48},
	year = {2015}}

@article{Je14,
	author = {Jeon, Jae-Hyung and Chechkin, Aleksei V. and Metzler, Ralf},
	doi = {10.1039/C4CP02019G},
	issue = {30},
	journal = {Physical Chemistry Chemical Physics},
	pages = {15811-15817},
	publisher = {The Royal Society of Chemistry},
	title = {Scaled Brownian motion: a paradoxical process with a time dependent diffusivity for the description of anomalous diffusion},
	url = {http://dx.doi.org/10.1039/C4CP02019G},
	volume = {16},
	year = {2014}}

@INPROCEEDINGS{Kac51,
       author = {{Kac}, M.},
        title = "{On Some Connections between Probability Theory and Differential and Integral Equations}",
    booktitle = {Second Berkeley Symposium on Mathematical Statistics and Probability},
         year = 1951,
       editor = {{Neyman}, Jerzy},
        month = jan,
        pages = {189-215},
       adsurl = {https://ui.adsabs.harvard.edu/abs/1951bsms.conf..189K}
}

@article{MeFlPa25,
  title = {Occupation-time statistics for non-Markovian random walks},
  author = {M\'endez, V. and Flaquer-Galm\'es, R. and Pal, A.},
  journal = {Physical Review E},
  volume = {111},
  issue = {4},
  pages = {044119},
  numpages = {20},
  year = {2025},
  month = {Apr},
  publisher = {American Physical Society},
  doi = {10.1103/PhysRevE.111.044119},
  url = {https://link.aps.org/doi/10.1103/PhysRevE.111.044119}
}

@article{BaFlMe23,
  title = {Ergodic properties of Brownian motion under stochastic resetting},
  author = {Barkai, E. and Flaquer-Galm\'es, R. and M\'endez, V.},
  journal = {Physical Review E},
  volume = {108},
  issue = {6},
  pages = {064102},
  numpages = {20},
  year = {2023},
  month = {Dec},
  publisher = {American Physical Society},
  doi = {10.1103/PhysRevE.108.064102},
  url = {https://link.aps.org/doi/10.1103/PhysRevE.108.064102}
}

@article{LeBa19,
  title = {Infinite ergodic theory for heterogeneous diffusion processes},
  author = {Leibovich, N. and Barkai, E.},
  journal = {Physical Review E},
  volume = {99},
  issue = {4},
  pages = {042138},
  numpages = {15},
  year = {2019},
  month = {Apr},
  publisher = {American Physical Society},
  doi = {10.1103/PhysRevE.99.042138},
  url = {https://link.aps.org/doi/10.1103/PhysRevE.99.042138}
}

@article{Si22,
  title = {Extreme value statistics and arcsine laws for heterogeneous diffusion processes},
  author = {Singh, Prashant},
  journal = {Physical Review E},
  volume = {105},
  issue = {2},
  pages = {024113},
  numpages = {12},
  year = {2022},
  month = {Feb},
  publisher = {American Physical Society},
  doi = {10.1103/PhysRevE.105.024113},
  url = {https://link.aps.org/doi/10.1103/PhysRevE.105.024113}
}

@misc{Yor92,
	author = {Yor, M},
	publisher = {Birkha user},
	title = {Some Aspects of Brownian Motion. Lectures in Math., ETH Z urich},
	year = {1992}
}

@article{Ge93,
	author = {Geman, H{\'e}lyette and Yor, Marc},
	doi = {https://doi.org/10.1111/j.1467-9965.1993.tb00092.x},
	journal = {Mathematical Finance},
	number = {4},
	pages = {349-375},
	title = {BESSEL PROCESSES, ASIAN OPTIONS, AND PERPETUITIES},
	volume = {3},
	year = {1993},
}

@article{Co05,
	author = {Comtet, Alain and Desbois, Jean and Texier, Christophe},
	doi = {10.1088/0305-4470/38/37/R01},
	journal = {Journal of Physics A: Mathematical and General},
	month = {aug},
	number = {37},
	pages = {R341},
	title = {Functionals of Brownian motion, localization and metric graphs},
	url = {https://dx.doi.org/10.1088/0305-4470/38/37/R01},
	volume = {38},
	year = {2005}}

@article{WeBa09,
  title = {Ergodic properties of fractional Brownian-Langevin motion},
  author = {Deng, Weihua and Barkai, Eli},
  journal = {Physical Review E},
  volume = {79},
  issue = {1},
  pages = {011112},
  numpages = {7},
  year = {2009},
  month = {Jan},
  publisher = {American Physical Society},
  doi = {10.1103/PhysRevE.79.011112},
  url = {https://link.aps.org/doi/10.1103/PhysRevE.79.011112}
}

@article{He08,
  title = {Random Time-Scale Invariant Diffusion and Transport Coefficients},
  author = {He, Y. and Burov, S. and Metzler, R. and Barkai, E.},
  journal = {Physical Review Letters},
  volume = {101},
  issue = {5},
  pages = {058101},
  numpages = {4},
  year = {2008},
  month = {Jul},
  publisher = {American Physical Society},
  doi = {10.1103/PhysRevLett.101.058101},
  url = {https://link.aps.org/doi/10.1103/PhysRevLett.101.058101}
}

@article{Th14,
  title = {Weak ergodicity breaking in an anomalous diffusion process of mixed origins},
  author = {Thiel, Felix and Sokolov, Igor M.},
  journal = {Physical Review E},
  volume = {89},
  issue = {1},
  pages = {012136},
  numpages = {7},
  year = {2014},
  month = {Jan},
  publisher = {American Physical Society},
  doi = {10.1103/PhysRevE.89.012136},
  url = {https://link.aps.org/doi/10.1103/PhysRevE.89.012136}
}

@article{Me15,
title = {A toolbox for determining subdiffusive mechanisms},
journal = {Physics Reports},
volume = {573},
pages = {1-29},
year = {2015},
issn = {0370-1573},
doi = {https://doi.org/10.1016/j.physrep.2015.01.002},
url = {https://www.sciencedirect.com/science/article/pii/S0370157315001404},
author = {Yasmine Meroz and Igor M. Sokolov}
}

@article{Bu10,
	author = {S. Burov and R. Metzler and E. Barkai},
	doi = {10.1073/pnas.1003693107},
	journal = {Proceedings of the National Academy of Sciences},
	number = {30},
	pages = {13228-13233},
	title = {Aging and nonergodicity beyond the Khinchin theorem},
	volume = {107},
	year = {2010}}

@article{Bau06,
	author = {A. Baule and R. Friedrich},
	doi = {https://doi.org/10.1016/j.physleta.2005.10.017},
	issn = {0375-9601},
	journal = {Physics Letters A},
	number = {3},
	pages = {167-173},
	title = {Investigation of a generalized Obukhov model for turbulence},
	url = {https://www.sciencedirect.com/science/article/pii/S0375960105015768},
	volume = {350},
	year = {2006}}

@article{MaBr02,
  title = {Large-deviation functions for nonlinear functionals of a Gaussian stationary Markov process},
  author = {Majumdar, Satya N. and Bray, Alan J.},
  journal = {Physical Review E},
  volume = {65},
  issue = {5},
  pages = {051112},
  numpages = {8},
  year = {2002},
  month = {May},
  publisher = {American Physical Society},
  doi = {10.1103/PhysRevE.65.051112},
  url = {https://link.aps.org/doi/10.1103/PhysRevE.65.051112}
}

@article{Ag84,
	author = {Agmon, Noam},
	doi = {10.1063/1.448113},
	issn = {0021-9606},
	journal = {The Journal of Chemical Physics},
	month = {10},
	number = {8},
	pages = {3644-3647},
	title = {Residence times in diffusion processes},
	url = {https://doi.org/10.1063/1.448113},
	volume = {81},
	year = {1984}}

@article{Ag10,
	author = {Noam Agmon},
	doi = {https://doi.org/10.1016/j.cplett.2010.08.019},
	issn = {0009-2614},
	journal = {Chemical Physics Letters},
	number = {4},
	pages = {184-186},
	title = {The residence time equation},
	url = {https://www.sciencedirect.com/science/article/pii/S0009261410011061},
	volume = {497},
	year = {2010}}

@article{MaCo02,
  title = {Local and Occupation Time of a Particle Diffusing in a Random Medium},
  author = {Majumdar, Satya  N. and Comtet, Alain},
  journal = {Physical Review Letters},
  volume = {89},
  issue = {6},
  pages = {060601},
  numpages = {4},
  year = {2002},
  month = {Jul},
  publisher = {American Physical Society},
  doi = {10.1103/PhysRevLett.89.060601},
  url = {https://link.aps.org/doi/10.1103/PhysRevLett.89.060601}
}

@article{SiKu19,
	author = {Singh, Prashant and Kundu, Anupam},
	doi = {10.1088/1742-5468/ab3283},
	journal = {Journal of Statistical Mechanics: Theory and Experiment},
	month = {aug},
	number = {8},
	pages = {083205},
	publisher = {IOP Publishing and SISSA},
	title = {Generalised `Arcsine' laws for run-and-tumble particle in one dimension},
	url = {https://dx.doi.org/10.1088/1742-5468/ab3283},
	volume = {2019},
	year = {2019}}

@article{Dh99,
  title = {Residence time distribution for a class of Gaussian Markov processes},
  author = {Dhar, Abhishek and Majumdar, Satya N.},
  journal = {Physical Review E},
  volume = {59},
  issue = {6},
  pages = {6413--6418},
  numpages = {0},
  year = {1999},
  month = {Jun},
  publisher = {American Physical Society},
  doi = {10.1103/PhysRevE.59.6413},
  url = {https://link.aps.org/doi/10.1103/PhysRevE.59.6413}
}

@book{von41,
  title={Invariant measures},
  author={Von Neumann, John},
  year={1941},
  publisher={American Mathematical Soc. Providence, Rhode Island}
}

@book{Kh49,
  title={Mathematical Foundations of Statistical Mechanics},
  author={Khinchin, A.I.},
  year={1949},
  publisher={Dover, New
York}
}

@book{vK92,
  title={Stochastic processes in physics and chemistry},
  author={Van Kampen, Nicolaas Godfried},
  volume={1},
  year={1992},
  publisher={Elsevier}
}

@article{Sa01,
	author = {Michael J. Saxton},
	doi = {https://doi.org/10.1016/S0006-3495(01)75870-5},
	issn = {0006-3495},
	journal = {Biophysical Journal},
	number = {4},
	pages = {2226-2240},
	title = {Anomalous Subdiffusion in Fluorescence Photobleaching Recovery: A Monte Carlo Study},
	url = {https://www.sciencedirect.com/science/article/pii/S0006349501758705},
	volume = {81},
	year = {2001}}

@article{Sz06,
	author = {Szyma{\'n}ski, J{\c e}drzej and Patkowski, Adam and Gapi{\'n}ski, Jacek and Wilk, Agnieszka and Ho{\l}yst, Robert},
	date = {2006/04/01},
	doi = {10.1021/jp055626w},
	isbn = {1520-6106},
	journal = {The Journal of Physical Chemistry B},
	journal1 = {The Journal of Physical Chemistry B},
	journal2 = {J. Phys. Chem. B},
	month = {04},
	number = {14},
	pages = {7367--7373},
	publisher = {American Chemical Society},
	title = {Movement of Proteins in an Environment Crowded by Surfactant Micelles:  Anomalous versus Normal Diffusion},
	type = {doi: 10.1021/jp055626w},
	url = {https://doi.org/10.1021/jp055626w},
	volume = {110},
	year = {2006},
	year1 = {2006}}

@article{Wu08,
	author = {Jianrong Wu and Keith M. Berland},
	doi = {https://doi.org/10.1529/biophysj.107.121608},
	issn = {0006-3495},
	journal = {Biophysical Journal},
	number = {4},
	pages = {2049-2052},
	title = {Propagators and Time-Dependent Diffusion Coefficients for Anomalous Diffusion},
	url = {https://www.sciencedirect.com/science/article/pii/S0006349508701620},
	volume = {95},
	year = {2008}}

@article{Ka23,
  title = {Anomalous diffusion and long-range memory in the scaled voter model},
  author = {Kazakevi\ifmmode \check{c}\else \v{c}\fi{}ius, Rytis and Kononovicius, Aleksejus},
  journal = {Physical Review E},
  volume = {107},
  issue = {2},
  pages = {024106},
  numpages = {16},
  year = {2023},
  month = {Feb},
  publisher = {American Physical Society},
  doi = {10.1103/PhysRevE.107.024106},
  url = {https://link.aps.org/doi/10.1103/PhysRevE.107.024106}
}

@article{Ri26,
  title={Atmospheric diffusion shown on a distance-neighbour graph},
  author={Richardson, Lewis Fry},
  journal={Proceedings of the Royal Society A},
  volume={110},
  number={756},
  pages={709--737},
  year={1926},
  publisher={The Royal Society London},
  doi     = {10.1098/rspa.1926.0043},
  url     = {https://doi.org/10.1098/rspa.1926.0043}
}

@article{Bo15,
	author = {Bodrova, Anna S and Chechkin, Aleksei V and Cherstvy, Andrey G and Metzler, Ralf},
	doi = {10.1088/1367-2630/17/6/063038},
	journal = {New Journal of Physics},
	month = {jun},
	number = {6},
	pages = {063038},
	publisher = {IOP Publishing},
	title = {Ultraslow scaled Brownian motion},
	url = {https://dx.doi.org/10.1088/1367-2630/17/6/063038},
	volume = {17},
	year = {2015}}

@article{Ko40,
  title={Wiener's spiral and some other interesting curves in Hilbert space},
  author={Kolmogorov, Andrei N},
  journal={Comptes Rendus de l'Académie des sciences de l'URSS},
  volume={26},
  pages={115--118},
  year={1940}
}

@article{Ma68,
  title={Fractional Brownian motions, fractional noises and applications},
  author={Mandelbrot, Benoit B and Van Ness, John W},
  journal={SIAM review},
  volume={10},
  number={4},
  pages={422--437},
  year={1968},
  publisher={SIAM},
  doi = {10.1137/1010093},
  URL = {https://doi.org/10.1137/1010093},
}

@article{Mi02,
  title={Is network traffic appriximated by stable L{\'e}vy motion or fractional Brownian motion?},
  author={Mikosch, Thomas and Resnick, Sidney and Rootz{\'e}n, Holger and Stegeman, Alwin},
  journal={The annals of applied probability},
  volume={12},
  number={1},
  pages={23--68},
  year={2002},
  publisher={Institute of Mathematical Statistics},
  doi={10.1214/aoap/1015961155}
}

@article{Co98,
  title={Long memory in continuous-time stochastic volatility models},
  author={Comte, Fabienne and Renault, Eric},
  journal={Mathematical Finance},
  volume={8},
  number={4},
  pages={291--323},
  year={1998},
  publisher={Wiley Online Library},
  doi = {https://doi.org/10.1111/1467-9965.00057}
}

@book{Bi08,
  title={Stochastic calculus for fractional Brownian motion and applications},
  author={Biagini, Francesca and Hu, Yaozhong and {\O}ksendal, Bernt and Zhang, Tusheng},
  year={2008},
  publisher={Springer}
}

@article{Je11,
  title={In vivo anomalous diffusion and weak ergodicity breaking of lipid granules},
  author={Jeon, Jae-Hyung and Tejedor, Vincent and Burov, Stas and Barkai, Eli and Selhuber-Unkel, Christine and Berg-S{\o}rensen, Kirstine and Oddershede, Lene and Metzler, Ralf},
  journal={Physical Review Letters},
  volume={106},
  number={4},
  pages={048103},
  year={2011},
  publisher={APS},
  doi = {10.1103/PhysRevLett.106.048103},
}

@article{Ta13,
  title={Intracellular transport of insulin granules is a subordinated random walk},
  author={Tabei, SM Ali and Burov, Stanislav and Kim, Hee Y and Kuznetsov, Andrey and Huynh, Toan and Jureller, Justin and Philipson, Louis H and Dinner, Aaron R and Scherer, Norbert F},
  journal={Proceedings of the National Academy of Sciences},
  volume={110},
  number={13},
  pages={4911--4916},
  year={2013},
  publisher={National Academy of Sciences},
  doi = {10.1073/pnas.1221962110},
}

@article{Je13,
  title={Anomalous diffusion and power-law relaxation of the time averaged mean squared displacement in worm-like micellar solutions},
  author={Jeon, Jae-Hyung and Leijnse, Natascha and Oddershede, Lene B and Metzler, Ralf},
  journal={New Journal of Physics},
  volume={15},
  number={4},
  pages={045011},
  year={2013},
  publisher={IOP Publishing},
  doi = {10.1088/1367-2630/15/4/045011},
}

@article{Je12,
  title={Anomalous diffusion of phospholipids and cholesterols in a lipid bilayer and its origins},
  author={Jeon, Jae-Hyung and Monne, Hector Martinez-Seara and Javanainen, Matti and Metzler, Ralf},
  journal={Physical Review Letters},
  volume={109},
  number={18},
  pages={188103},
  year={2012},
  publisher={APS},
  doi = {10.1103/PhysRevLett.109.188103},
}

@article{Ma09,
  title={Fractional Brownian Motion Versus the Continuous-Time Random Walk: A Simple Test for Subdiffusive Dynamics},
  author={Magdziarz, Marcin and Weron, Aleksander and Burnecki, Krzysztof and Klafter, Joseph},
  journal={Physical Review Letters},
  volume={103},
  number={18},
  pages={180602},
  year={2009},
  publisher={APS},
  doi = {10.1103/PhysRevLett.103.180602},
}

@article{Re15,
  title={Superdiffusion dominates intracellular particle motion in the supercrowded cytoplasm of pathogenic Acanthamoeba castellanii},
  author={Reverey, Julia F and Jeon, Jae-Hyung and Bao, Han and Leippe, Matthias and Metzler, Ralf and Selhuber-Unkel, Christine},
  journal={Scientific reports},
  volume={5},
  number={1},
  pages={11690},
  year={2015},
  publisher={Nature Publishing Group UK London},
  doi={https://doi.org/10.1038/srep11690}
}

@article{Kr19,
  title={Spectral content of a single non-Brownian trajectory},
  author={Krapf, Diego and Lukat, Nils and Marinari, Enzo and Metzler, Ralf and Oshanin, Gleb and Selhuber-Unkel, Christine and Squarcini, Alessio and Stadler, Lorenz and Weiss, Matthias and Xu, Xinran},
  journal={Physical Review X},
  volume={9},
  number={1},
  pages={011019},
  year={2019},
  publisher={APS},
  doi = {10.1103/PhysRevX.9.011019},

}

@book{Qi08,
  title={Processes with long-range correlations: Theory and applications},
  author={Rangarajan, Govindan and Ding, Mingzhou},
  volume={621},
  year={2008},
  publisher={Springer}
}

@article{DeBa09,
	author = {Deng, Weihua and Barkai, Eli},
	journal = {Physical Review E},
	number = {1},
	pages = {011112},
	publisher = {APS},
	title = {Ergodic properties of fractional Brownian-Langevin motion},
	volume = {79},
	year = {2009},
    doi = {10.1103/PhysRevE.79.011112},

}

@article{Sa18,
	author = {Sadhu, Tridib and Delorme, Mathieu and Wiese, Kay J{\"o}rg},
	journal = {Physical Review Letters},
	number = {4},
	pages = {040603},
	publisher = {APS},
	title = {Generalized arcsine laws for fractional Brownian motion},
	volume = {120},
	year = {2018},
    doi = {10.1103/PhysRevLett.120.040603},
}

@article{Ki22,
	author = {Kimura, Mutsumi and Akimoto, Takuma},
	journal = {Physical Review E},
	number = {6},
	pages = {064132},
	publisher = {APS},
	title = {Occupation time statistics of the fractional Brownian motion in a finite domain},
	volume = {106},
	year = {2022},
    doi = {10.1103/PhysRevE.106.064132},
    url = {https://link.aps.org/doi/10.1103/PhysRevE.106.064132}
}

@article{Je10,
	author = {Jeon, Jae-Hyung and Metzler, Ralf},
	journal = {Physical Review E},
	number = {2},
	pages = {021103},
	publisher = {APS},
	title = {Fractional Brownian motion and motion governed by the fractional Langevin equation in confined geometries},
	volume = {81},
	year = {2010},
    doi = {10.1103/PhysRevE.81.021103},

}

@article{Vo21,
  title={Probability density of fractional Brownian motion and the fractional Langevin equation with absorbing walls},
  author={Vojta, Thomas and Warhover, Alex},
  journal={Journal of Statistical Mechanics: Theory and Experiment},
  volume={2021},
  number={3},
  pages={033215},
  year={2021},
  publisher={IOP Publishing},
  doi={10.1088/1742-5468/abe700}
}

@article{Wi11,
  title = {Perturbation theory for fractional Brownian motion in presence of absorbing boundaries},
  author = {Wiese, Kay J\"org and Majumdar, Satya N. and Rosso, Alberto},
  journal = {Physical Review E},
  volume = {83},
  issue = {6},
  pages = {061141},
  numpages = {15},
  year = {2011},
  month = {Jun},
  publisher = {American Physical Society},
  doi = {10.1103/PhysRevE.83.061141},
  url = {https://link.aps.org/doi/10.1103/PhysRevE.83.061141}
}

@article{Safd2015,
  title = {Aging scaled Brownian motion},
  author = {Safdari, Hadiseh and Chechkin, Aleksei V. and Jafari, Gholamreza R. and Metzler, Ralf},
  journal = {Physical Review E},
  volume = {91},
  issue = {4},
  pages = {042107},
  numpages = {9},
  year = {2015},
  month = {Apr},
  publisher = {American Physical Society},
  doi = {10.1103/PhysRevE.91.042107},
  url = {https://link.aps.org/doi/10.1103/PhysRevE.91.042107}
}

@article{Darling_1957,
    title={On occupation times for Markoff processes},
    year={1957},
    author={D. A. Darling and Mark Kac},
    doi={10.1090/s0002-9947-1957-0084222-7},
    mag_id={2008594757},
    journal={Transactions of the American Mathematical Society},
    volume={84},
    pages={444-458}
}

@article{Wa22,
	author = {Wang, Wei and Metzler, Ralf and Cherstvy, Andrey G.},
	doi = {10.1039/D2CP01741E},
	issue = {31},
	journal = {Physical Chemistry Chemical Physics},
	pages = {18482-18504},
	publisher = {The Royal Society of Chemistry},
	title = {Anomalous diffusion{,} aging{,} and nonergodicity of scaled Brownian motion with fractional Gaussian noise: overview of related experimental observations and models},
	url = {http://dx.doi.org/10.1039/D2CP01741E},
	volume = {24},
	year = {2022}}

@ARTICLE{CBM-fBM,
  author={Perrin, E. and Harba, R. and Jennane, R. and Iribarren, I.},
  journal={IEEE Signal Processing Letters}, 
  title={Fast and exact synthesis for 1-D fractional Brownian motion and fractional Gaussian noises}, 
  year={2002},
  volume={9},
  number={11},
  pages={382-384},
  doi={10.1109/LSP.2002.805311}
}

\end{document}